\documentclass[11pt]{article}
\usepackage{epsfig,dsfont}
\usepackage{geometry}
\begin{document}  
\sffamily

\thispagestyle{empty}
\vspace*{15mm}

\begin{center}

{\huge
Simulation strategies for the massless lattice 
\vskip3mm
Schwinger model in the dual formulation}
\vskip25mm
Daniel G\"oschl$^{a}$, Christof Gattringer$^{a}$, Alexander Lehmann$^{b}$, Christoph Weis$^{c}$ 
\vskip10mm
$^a$ Universit\"at Graz, Institut f\"ur Physik, 8010 Graz, Austria \\
$^b$ Universit\"at Heidelberg,
Institut f\"ur Theoretische Physik, 69120 Heidelberg, Germany \\
$^c$ University of Oxford, Mathematical Institute, OX2 6GG Oxford, United Kingdom
\end{center}
\vskip30mm

\begin{abstract}
The dual form of the massless Schwinger model on the lattice overcomes the complex action 
problems from two sources: a topological term, as well as non-zero chemical potential, making these 
physically interesting cases accessible to Monte Carlo simulations. The partition function is represented as a sum 
over fermion loops, dimers and plaquette-surfaces such that all contributions are real and positive. However,
these new variables constitute a highly constrained system and suitable update strategies have to be developed. 
In this exploratory study we present an approach based on locally growing plaquette-surfaces surrounded by fermion loop 
segments combined with a worm based strategy for updating chains of dimers, as well as winding fermion loops. 
The update strategy is checked with conventional simulations as well as reference data from exact summation on small 
volumes and we discuss some physical implications of the results.  
\end{abstract}

\newpage
\setcounter{page}{1}

\section{Introduction}
Calculations at finite density are considered to be one of the great challenges for lattice field theory Monte Carlo simulations.
The quest for an ab-initio understanding of non perturbative features of quantum field theories, such as phase diagrams at 
non-vanishing chemical potentials, have given rise to a considerable number of innovative approaches to the problem and  
reviews at the yearly lattice conferences \cite{reviews} provide an overview about ideas that were pursued in recent years. 
At the core of the challenge is the so called ''complex action problem'' or ''sign problem'': 
Many theories have a complex action $S$ when a chemical
potential, or also a topological term (theta term, vacuum term) are coupled. 
In these cases the Boltzmann factor $\exp( -S)$ in the path integral
has a complex phase and cannot be interpreted as a probability in a Monte Carlo simulation.

An elegant approach that entirely overcomes the complex action problem
is to rewrite a lattice field theory in terms of new degrees of freedom, often referred to 
as ''dual variables''. In a successful dualization the theory is exactly rewritten in terms of the dual variables such 
that in the new representation the partition sum has only real and positive
contributions. The dual variables are world-lines for matter fields and world-sheets for gauge fields. The 
world-sheets can either form closed surfaces or surfaces with a boundary formed by a matter loop. For bosonic abelian 
theories an impressive body of work based on the dual approach was presented in recent years 
(for examples see \cite{abelian,scalarqed2}). 

For relativistic fermions the situation is complicated further by additional signs from the Grassmann nature of the fermion fields
and the commutators of the Dirac matrices. 
Thus only relatively few real and positive dual representations for systems with fermions can be found
in the literature, often in the limit of strong coupling or the sign quenched approximation 
(for some examples see, e.g., \cite{fermions,schwinger1, schwinger2, schwinger3}). 
A case where a real and positive representation is known for arbitrary couplings is 2-dimensional QED, i.e., the Schwinger model, 
with chemical potential and a topological term \cite{schwinger_dual}. Both, the chemical potential, as well as the topological term 
lead to complex action problems and the dual representation in terms of world-lines, dimers and world-sheets given in 
\cite{schwinger_dual} solves both these problems in principle by providing an exact representation where all contributions to the 
partition sum are real and positive. 

In this exploratory paper we present strategies for simulating the dual representation \cite{schwinger_dual}. More specifically we 
consider the following two cases: 1) the one flavor case with a topological term (at vanishing chemical potential) and, 2) the case of
two oppositely charged flavors with a non-zero chemical potential (no topological term coupled). 
In both cases one has to update the fermion loops 
together with the world-sheets that are bounded by the world-lines. In addition one has to sample the dimer 
contributions from the fermion integral which do not couple to the gauge degrees of freedom. The fermion loops and dimers 
constitute a highly constrained system and here we propose and test strategies to update them in accordance with the constraints 
coming from the gauge symmetries of the conventional representation. We test our algorithms against exact calculations 
on small lattices and conventional simulations without topological term or chemical potential. The results presented here constitute the first 
simulation of the Schwinger model at arbitrary, i.e., weak couplings with a finite chemical potential and a topological term.

We expect that the explorative study of simulation strategies presented here will be useful for simulating the highly 
constrained world-line/world-sheet 
representations of other fermionic systems. Furthermore, the reference data for the Schwinger model which can be obtained 
from the dual approach tested here, will be useful also for assessing other techniques that are explored
for overcoming complex action problems: In particular methods based on complexification 
(see, e.g., the reviews \cite{thimblecl}) or new strategies for simulating systems at finite vacuum term 
\cite{thetasimulations} can be cross-checked against results from the world-line/world-sheet methods presented here.

\section{The models and their dual representation}
As announced in the introduction, in this paper we study two variants of the massless Schwinger model on the lattice: 
the one flavor case with a topological term, and the case of two oppositely charged fermions with chemical potentials. 
This section serves to present the two models in their conventional representation and to summarize the dual formulation 
presented in \cite{schwinger_dual}, which is the basis for the dual simulation discussed here.

\subsection{The one flavor model with topological term}
For the one flavor case the partition sum is given by
\begin{equation}
\label{eq_conv_part}
Z  \; = \; \int \! \mathcal{D}[U]\mathcal{D}\big[\,\overline{\psi},\psi \,\big]~e^{-\, S_G[U]  \, -i\,\theta \, Q[U] \, - \, 
S_\psi[U,\overline{\psi},\psi] }  \; ,
\end{equation}
where $U_\nu(n) = \exp(i A_\nu(n)), \, A_\nu(n) \in [-\pi, \pi]$ are the U(1)-valued link variables for the gauge fields. 
By $n = (n_1,n_2)$ we denote the sites of a 2-dimensional $N_S \times N_T$ lattice and for a technical simplification in the 
dual representation (see \cite{schwinger_dual}) we restrict ourselves to lattices where $N_T$ and $N_S$ are multiples
of $4$. The index $\nu = 1,2$ runs over the Euclidean space ($\nu = 1$) and time ($\nu = 2$) directions, and
$V= N_S N_T$ is used to denote the total number of lattice points. For the fermions we use one-component Grassmann valued 
variables $\overline{\psi}(n)$ and $\psi(n)$. In the conventional representation (\ref{eq_conv_part}) 
all boundary conditions are periodic, 
except for the temporal boundary conditions of the fermions which have to be chosen anti-periodic. The topological 
charge $Q[U]$ is coupled with the vacuum angle $\theta$.

In the path integral the gauge links are integrated with the product measures $\mathcal{D}[U]$
of U(1) Haar-measures and the fermions with the Grassmann product measure $\mathcal{D}\big[\,\overline{\psi},\psi \,\big]$,
\begin{equation}
\label{measures}
\int \! \mathcal{D}[U]  \; = \; \prod_{n,\nu} \int_{-\pi}^\pi \! \frac{d A_\nu (n)}{2 \pi} \; \; \; , \; \; \;  \; \;
\int \! \mathcal{D}\big[\, \overline{\psi},\psi \, \big] \; = \; \prod_{n} \! \int d\overline{\psi}(n) \, d \psi(n) \; .
\end{equation}
For the gauge action $S_G[U]$ we use the Wilson form,
\begin{equation}
\label{eq_conv_gauge}
S_G[U] \; = \; -\beta \sum_{n} \mbox{Re} \, U_p(n)  \; = \; -\frac{\beta}{2} \sum_{n} \big[ \, U_p(n) \, + \, U_p(n)^{\star} \, \big] \; ,
\end{equation}
where the plaquettes $U_p(n)$, labelled by the coordinate $n$ 
of their lower left corner, are the products $U_p(n) \, = \,U_1(n) \, U_2(n+\hat{1}) \, U_1 (n+\hat{2})^{\star} \, U_2(n)^{\star}$
and $\beta$ is the inverse gauge coupling.
The topological charge $Q[U]$ is introduced using the field theoretical definition 
\begin{equation}
\label{eq_conv_topcharge}
Q[U] \; = \; \frac{1}{i 4 \pi} \sum_n \big[ \, U_p(n) \, - \, U_p(n)^{\star} \, \big] \; ,
\end{equation}
which in the continuum limit ($\beta \rightarrow \infty$) goes over into the integer valued topological charge
of the continuum. The massless staggered fermions are described by the action
\begin{equation}
\label{psiaction}
S_\psi[U,\overline{\psi},\psi] \; = \; \frac{1}{2} \sum_{n, \nu } \gamma_\nu(n) \,\Big[  U_\nu(n) \, \overline{\psi}(n) \, \psi(n +\hat{\nu}) \; - \; 
 U_\nu(n)^{\star} \, \overline{\psi}(n + \hat{\nu}) \, \psi(n) \Big] \; ,
\end{equation}
with the staggered sign function $\gamma_\nu(n)$ given by
$\gamma_1(n)  = 1 \, , \;  \gamma_2(n) = (-1)^{n_1}$.
We can write the partition function as
\begin{equation}
\label{partsum}
Z  \; = \; \int \!\! \mathcal{D}[U]\mathcal{D}\big[ \, \overline{\psi},\psi \, \big] \;  \; 
e^{ \, \eta \sum_n U_p(n)} \; \; e^{ \, \overline\eta \sum_n U_p(n)^{\star}} \; 
e^{  - \, S_\psi[U,\overline{\psi},\psi] } \; ,
\end{equation}
where we have introduced the abbreviations
$\eta  = \frac{\beta}{2}-\frac{\theta}{4\pi}\, , \; \overline{\eta} = \frac{\beta}{2}+\frac{\theta}{4\pi}$.
The conventional representation (\ref{partsum}) has a complex action problem for non-zero values of the vacuum angle 
\(\theta\), since then 
$\eta \neq \overline\eta\,$, such that the Boltzmann factor has a complex phase and cannot be used as a probability weight 
in a Monte Carlo update. We remark, that the physically interesting continuum limit is reached for $\beta \rightarrow 
\infty$ where both parameters $\eta$ and $\overline{\eta}$ are positive. As we will see below
it is sufficient to restrict $\beta$ to values $\beta > |\theta|/2 \pi$, which ensures a particularly simple form of the 
dual representation.

\begin{figure}[t!]
\begin{center}
\includegraphics[width=6.8cm,type=pdf,ext=.pdf,read=.pdf]{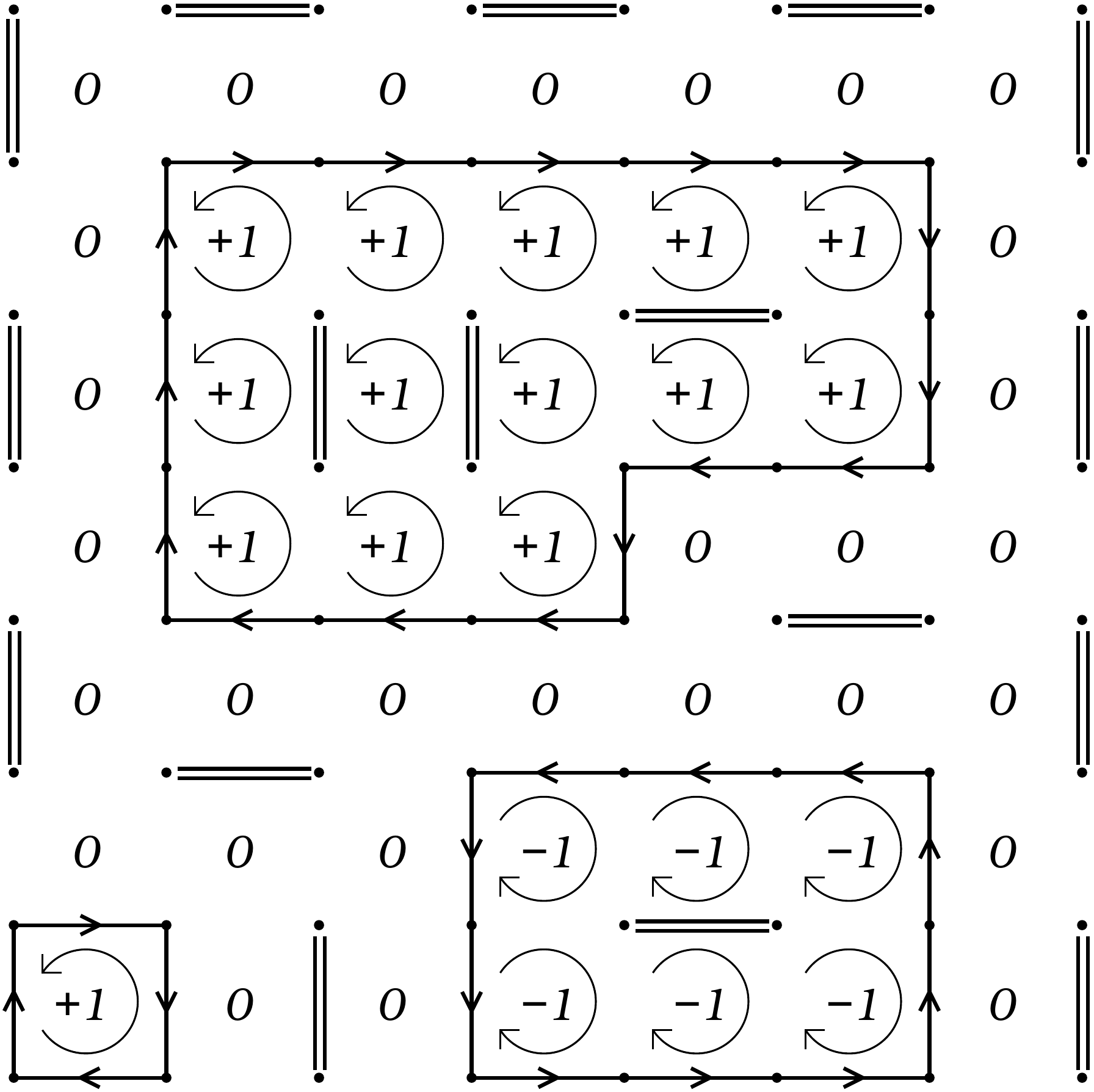}
\end{center}
\caption{Example of an admissible configuration of the one flavor model 
on an $8 \times 8$ lattice (from \cite{schwinger_dual}). The lattice is filled 
with oriented self-avoiding loops (single lines with arrows) and dimers (double lines), such that each site is either run 
through by a loop or is the endpoint of a dimer. Flux introduced by loops has to be compensated by placing 
plaquette occupation numbers, which are specified by the integers we write inside each plaquette.}
\label{fig1}
\end{figure}

To overcome the complex action problem an exact rewriting of the partition function 
in terms of new degrees of freedom, so-called dual variables, was presented in \cite{schwinger_dual}.
In this dual representation each term in the partition sum 
is real and positive, such that one can access finite values of $\theta$ with Monte Carlo simulations. The transformation 
is based on an expansion of the Boltzmann factors and a subsequent exact integration of the original field variables.
The dual form of the partition sum is given by (compared to the form given in \cite{schwinger_dual} 
we have dropped an overall irrelevant factor of $2^{-V}$)
\begin{equation}
Z \; = \;   \sum_{\{l,d,p\}}  
\prod_n I_{|p(n)|}\!\left( 2 \sqrt{\eta \overline{\eta}} \, \right) \, \left( \sqrt{\frac{\eta}{\overline{\eta}}} \, \right)^{p(n)} \; .
\label{zfinal2}
\end{equation}
The sum $\sum_{\{l,d,p\}}$ in (\ref{zfinal2}) runs over all admissible configurations of the dual variables,
non-intersecting oriented fermion loops $l$, dimers $d$ and plaquette occupation numbers $p(n)$. 
A configuration is admissible when each site of the lattice is either run through by a fermion loop or is the endpoint
of a dimer. The fermion loops introduce fluxes along their contour. These fluxes have to be compensated by filling the fermion loop 
with plaquette occupation numbers $p(n) \in \mathds{Z}$. A value $p(n) = +k \, ($or $-k)$ with $k > 0$ gives rise to $k$ 
units of flux around the plaquette at site $n$ with mathematically positive (negative) orientation. If two neighboring plaquettes are not 
separated  by a segment of fermion loop, then they must have the same plaquette occupation numbers, such that the 
flux along the link where they touch is compensated. In the dual representation all boundary conditions are periodic.
An example of an admissible configuration on an $8 \times 8$ 
lattice is shown in Fig.~\ref{fig1}. Note that we could also increase or decrease all plaquette occupation 
numbers by the same integer and still have an admissible configuration.

Each configuration $(l,d,p)$ comes with a weight factor 
$\prod_n I_{|p(n)|} ( 2 \sqrt{\eta \overline{\eta}} ) \, ( \sqrt{ \eta / \overline{\eta} } \,)^{p(n)}$ 
that depends only on the plaquette occupation numbers $p(n)$ which are restricted by the admissibility
condition of the configuration. By $I_p$ we denote the modified Bessel functions, and obviously the 
weights are real and positive (assuming that we choose $\beta > |\theta|/2 \pi$ such that $\eta$ and $\overline{\eta}$ are positive). 
Thus the dual representation solves the complex action problem completely. 

Note that a non-zero vacuum angle $\theta$ gives different weights to positive and negative plaquette 
occupation numbers via the factor $( \sqrt{ \eta / \overline{\eta}} \,)^{p(n)}$. 
For $\theta > 0$ we have $\eta < \overline{\eta}$ and thus $\sqrt{\eta/\overline{\eta}} < 1$, which implies that for $\theta > 0$, 
negative plaquette occupation numbers $p(n)$ have a larger weight  and thus are favored (and vice versa).

In this study we focus on bulk observables which can be obtained as derivatives of $\ln Z$. In particular we compute 
the plaquette expectation
value $\langle U_p \rangle$ and its susceptibility $\chi_p$, as well as the topological charge density $\langle q \rangle$ and the 
topological susceptibility $\chi_t$,
\begin{eqnarray}
\langle U_p \rangle & = & \frac{1}{V} \frac{\partial}{\partial \beta} \ln Z \quad \; , \qquad 
\chi_p \; = \; \frac{1}{V} \frac{\partial^2}{\partial \beta^2} \ln Z \; ,
\label{plaqvev} \\ 
\langle q \rangle& = &  \frac{-1}{V} \frac{\partial}{\partial \theta} \ln Z \quad , \qquad 
\chi_t \; = \; \frac{1}{V} \frac{\partial^2}{\partial \theta^2} \ln Z \; .
\label{topovev}
\end{eqnarray}
These derivatives can easily be evaluated also in the dual representation and give rise to weighted sums of the modified 
Bessel functions and their moments.

\subsection{Two flavor model with chemical potential}

The conventional representation of the lattice partition sum for the two flavor model with chemical potential reads
(we omit the topological term here) 
\begin{equation}
\label{eq_conv_part_2_f}
Z  \; = \; \int \! \mathcal{D}[U] \, \mathcal{D}[\, \overline{\psi},\psi] \, \mathcal{D}[\overline{\chi},\chi] \,
e^{\,-\, S_G[U]  \,  - \, S_{\psi}[U,\overline{\psi},\psi]
- \, S_{\chi}[U,\overline{\chi},\chi] }  \; .
\end{equation}
$\psi$, $\overline{\psi}$ and $\chi$, $\overline{\chi}$ are two fermion flavors with opposite charge. 
Like the measure $\mathcal{D}[\,\overline{\psi},\psi]$ given in (\ref{measures}), also the measure $\mathcal{D}[\overline{\chi},\chi]$ 
for the second flavor $\chi$ is a product over Grassmann measures at all sites. The fermionic action now contains 
chemical potentials $\mu_\psi$ and $\mu_\chi$ for both flavors:
\begin{equation}
\label{chimuaction}
S_{\psi}[U,\overline{\psi},\psi] \; = \; \frac{1}{2} \sum_{n, \nu } \gamma_\nu(n) \Big[  e^{\mu_\psi \delta_{\nu,2}} \, U_\nu(n) \, 
\overline{\psi}(n) \, \psi(n \!+\!\hat{\nu}) \; - \;  e^{-\mu_\psi \delta_{\nu,2}}\, U_\nu(n)^{*} \, \overline{\psi}(n \!+ \! \hat{\nu}) \, \psi(n) \Big] ,
\end{equation}
\begin{equation}
\label{chimuaction2}
S_{\chi}[U,\overline{\chi},\chi] \; = \; \frac{1}{2} \sum_{n, \nu } \gamma_\nu(n) \Big[ 
e^{\mu_\chi \delta_{\nu,2}} \, U_\nu(n)^{*} \, \overline{\chi}(n) \, \chi(n \!+\!\hat{\nu}) \; - \; 
e^{-\mu_\chi \delta_{\nu,2}}\, U_\nu(n) \, \overline{\chi}(n \!+ \!\hat{\nu}) \, \chi(n) \Big] .
\end{equation}
Since the second flavor $\chi$ has negative charge, in $S_{\chi}[U,\overline{\chi},\chi]$
the link variables $U_\nu(n)$ are interchanged with their 
complex conjugate $U_\nu(n)^{*}$. This ensures overall electric neutrality of the two flavor 
system as required by Gauss' law. The chemical potentials $\mu_\psi$ and $\mu_\chi$ for the two flavors
give different weights to temporal ($\nu = 2$) forward and backward hops.

We remark that for the two flavor case considered here physics will only depend on the sum of the 
two chemical potentials $\mu_\psi + \mu_\chi$ due to Gauss' law. However, for more than two flavors 
(keeping overall electric neutrality) physics will depend also on non-trival combinations of 
the individual chemical potentials (see, e.g., \cite{narayanan}). The dualization for more than 
two flavors is straightforward \cite{schwinger_dual}, and with this possible generalization in mind 
we find it instructive to explicitly show the dependence on the individual chemical potentials.
 
For the dual representation of the two flavor model we now have a second set of dual variables, 
$\overline{l}, \overline{d}$ for the fermion loops and the
dimers of the second fermion flavor. Together with $l$ and $d$ for the first flavor and the plaquette occupation numbers $p(n)$ 
this constitutes the set of dual variables of the two flavor case. The dual form of the partition function of the two flavor case 
is then given as a sum over the configurations of all dual variables (compared to \cite{schwinger_dual} 
we again drop an irrelevant overall factor)
\begin{equation}
Z \; = \;  \sum_{\{l,d,\overline{l},\overline{d},p\}} e^{\, \mu_\psi \, N_T W(l)} \, 
e^{\, \mu_\chi \, N_T W(\overline{l})} 
 \, \prod_n I_{|p(n)|}\!\left( \beta \right) \; .
\label{z2flavordual}
\end{equation}
In an admissible configuration the fermion constraints have to be obeyed for both flavors, i.e., for both, the 
$l,d$ and the $\overline{l}, \overline{d}$ variables each site has to be either the endpoint of a dimer or run through by a fermion loop.
At each link the combined flux of the fermion loops from both flavors has to be compensated by activated plaquettes. Since 
the second flavor has negative charge the flux from the $\overline{l}$-loops is counted with a negative sign, i.e., here 
the flux from the loop $\overline{l}$ and from the plaquettes $p(n)$ have to be equal and run in the same direction for cancellation.
Also equal fluxes from $l$ and $\overline{l}$ along a link that run in the same direction saturate the gauge constraints on that link. 
The rhs.\ of Fig.~\ref{admissible_configuration_2_flavors} shows such a double fermion loop (we use dashed lines for
the $\overline{l}$ loops and the $\overline{d}$ dimers). 

The two fermion flavors interact with each other only on those plaquettes which are connected to fermion 
loops of both flavors, such that the weight
$I_{|p(n)|}(\beta)$ has a different $p(n)$ from what it would have for only a fermion loop of one flavor. 
We remark that also in the two flavor model
one can add the topological term, simply be replacing $I_{|p(n)|}(\beta)$ with 
$I_{|p(n)|} ( 2\,\sqrt{\eta \overline{\eta}} \, ) \; ( \sqrt{\eta / \overline{\eta}} \, )^{p(n)}$ 
(see \cite{schwinger_dual}). However, in (\ref{z2flavordual}) we use the simpler dual form for $\theta = 0$, since this 
is the case we actually simulate. 

In the dual representation the chemical potentials $\mu_\psi$ and $\mu_\chi$ couple to the total temporal winding numbers $W(l)$ 
and $W(\overline{l})$ of the loops $l$ and $\overline{l}$ that correspond to the flavors $\psi$ and $\chi$. 
In Fig.~\ref{admissible_configuration_2_flavors} the double fermion loop on the rhs.\ of the plot couples to the chemical potential 
with the factor $e^{(\mu_\psi + \mu_\chi) \, N_T\!}$.
It is obvious, that the weights in the partition sum (\ref{z2flavordual}) are real and positive also for arbitrary values of the chemical 
potentials $\mu_\psi$ and $\mu_\chi$. Thus in the dual formulation the sign problem is gone completely also for 
the case of finite chemical potential. 

\begin{figure}[t]
\centering
	\centering
	\includegraphics[scale=0.28]{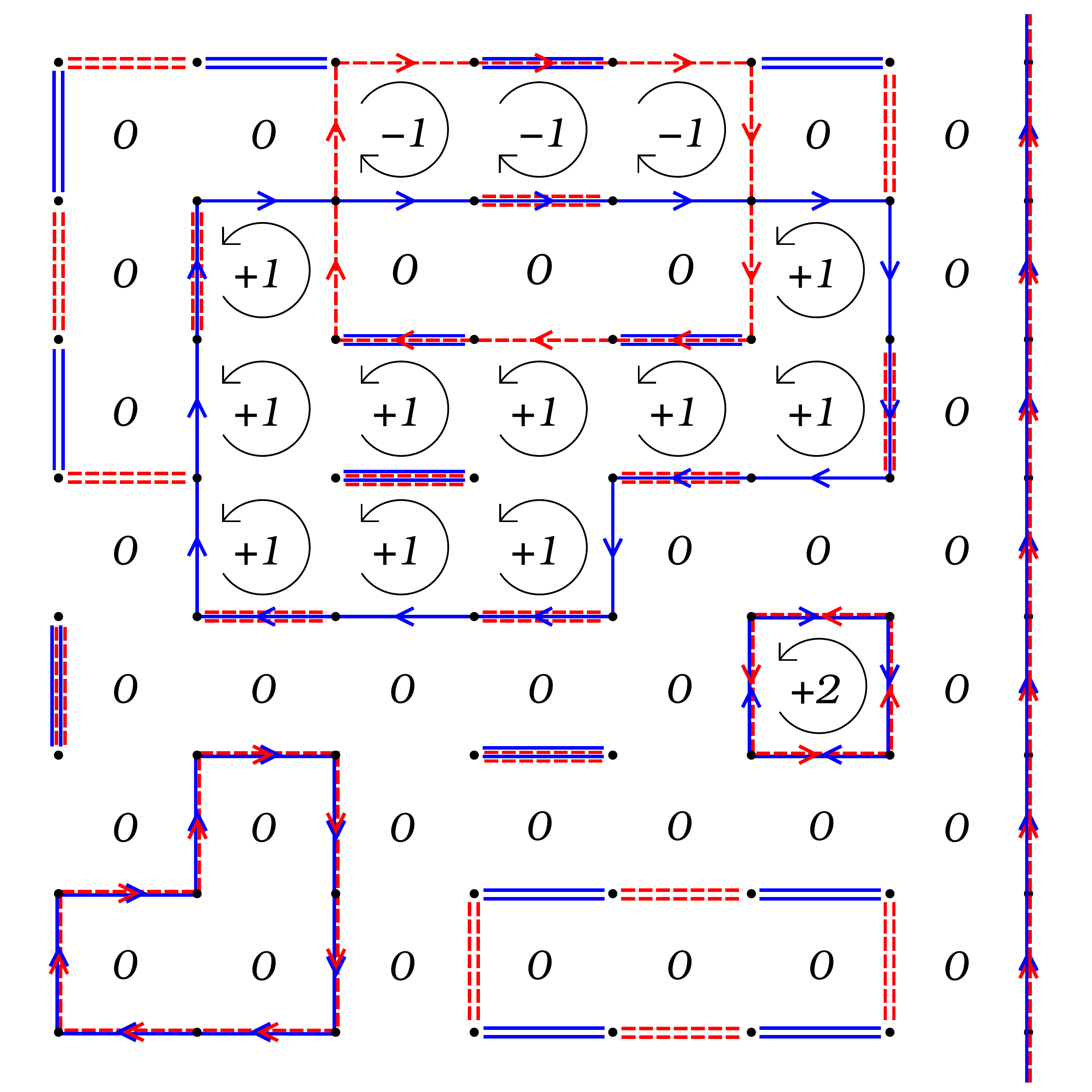}
	\caption{Example of an admissible configuration for the two flavor model on a $8\times 8$ lattice. Again we use 
	double lines to represent dimers and single lines with arrows for fermion 
	loops. However, here we have fermion loops and dimers for both 
	flavors and in order to distinguish them in the figure we use dashed lines for the loops and dimers of the second flavor.
	Fermion loops have to be filled with activated plaquettes (again represented by circular arrows with the occupation 
	number written inside) such that along the loop the flux is saturated and the gauge constraints are obeyed. Note that 
	for the second flavor (dashed lines) which is negatively charged, the gauge flux and the loop must run 
	in the same direction for saturation. This is the reason why also loops from different flavors that run parallel to each other satisfy 
	the gauge constraints.  Finally we remark, that the vertical double loop on the rhs.\ of the plot is an example of a loop that couples
	to the chemical potential (both flavors have temporal winding number $W(l) = W(\overline{l}) = +1$ in that configuration).}
	\label{admissible_configuration_2_flavors}
\end{figure}

In addition to the plaquette expectation number and its susceptibility defined in (\ref{plaqvev}), in the two flavor model with 
chemical potential we can also study the particle number densities $n_{\psi}$, $n_{\chi}$ and the corresponding susceptibilities
$\chi_{{\psi}}$, $\chi_{{\chi}}$. They are defined as 
\begin{equation}
\langle n_{\psi/\chi} \rangle \; = \; \frac{1}{N_S} \frac{\partial \ln (Z)}{\partial (\mu_{\psi/\chi} N_T)} \quad , \; \quad
\chi_{{\psi/\chi}} \; = \; \frac{1}{N_S} \frac{\partial^2 \ln (Z)}{\partial (\mu_{\psi/\chi} N_T)^2} \; .
\end{equation}
Again we can apply the derivatives also to the dual form of the model to obtain the dual representation of these observables. 
Their dual form is particularly simple,
\begin{eqnarray}
\langle n_\psi \rangle &=& \frac{1}{N_S} \langle W(l) \rangle  \quad , \; \quad
\chi_{\psi} \; = \; \frac{1}{N_S} \left[ \langle W(l)^2 \rangle - \langle W(l) \rangle^2 \right] \; , 
\nonumber \\	
\langle n_\chi \rangle &=& \frac{1}{N_S} \langle W(\bar{l}) \rangle  \quad , \; \quad
\chi_{\chi}  \; = \; \frac{1}{N_S} \left[ \langle W(\bar{l})^2 \rangle - \langle W(\bar{l}) \rangle^2 \right] \ ,
\end{eqnarray}
i.e., the particle densities and susceptibilities 
are the first and second (connected) moments of the corresponding temporal winding numbers $W(l)$ and $W(\bar{l})$. 
These expressions illustrate an 
elegant feature of the dual representation: the net particle number can be identified as an integer on every single configuration. 
This is not the case for the conventional formulation, where the particle number cannot be defined as an integer on 
a single configuration. The fact that the particle number can be uniquely identified in the dual representation opens the door
to canonical simulations, i.e., simulations at fixed particle number, which we briefly discuss at the end of 
Section \ref{twoparticle}  where we consider the two flavor case.

\section{Update strategies and results for the one flavor model}
\label{oneparticle}

We begin the discussion of the dual update strategies with the one flavor model with a topological term, since the one flavor
case is  simpler and the two flavor updates build on the steps developed here. 

\subsection{Steps of the update}

The partition sum (\ref{zfinal2}) is a sum over configurations of oriented loops and dimers. These have to obey the fermion 
constraints, i.e., every site of the lattice is either the endpoint of a dimer or is run through by a loop. The weight of each 
configuration, $\prod_n I_{|p(n)|}\!\left( 2 \sqrt{\eta \overline{\eta}} \, \right) \, ( \sqrt{ \eta / \overline{\eta}} \, )^{p(n)}$,
is computed from the plaquette occupation numbers $p(n)$, which have to be such that the gauge constraints for the links along 
the fermion loops are obeyed (compare the example in Fig.~\ref{fig1}). 

Let us begin with the update of the dimers in a given fixed configuration of fermion loops. It is easy to see from the example shown in 
Fig.~\ref{fig1} that different configurations of the dimers are compatible with a given configuration of loops. 
For example the two vertical dimers inside the loop at the top of the configuration can be replaced by a pair of horizontal dimers. 
Another possible modification is to shift the vertical line of dimers on the rhs.\ of the configuration upwards by one link (note
that the lattice closes periodically). More generally one can identify closed contours where dimers alternate with empty links and shift
all dimers along that contour by one link. This is a move that modifies only the dimers and 
leaves all fermion constraints intact without altering the fermion loops. 
Furthermore these moves of the dimers do not change the weights, since all plaquette occupation numbers $p(n)$ remain the 
same and no Metropolis decision is needed for accepting a pure dimer change. It has to be pointed out 
that some dimers are frozen by the fermion 
loops surrounding them, e.g., the horizontal dimer in the top fermion 
loop, or the dimer inside the $2\times 3$ fermion loop at the bottom. 

The idea of identifying closed contours where dimers and empty links alternate naturally leads to a worm update. We start the worm 
from a randomly chosen lattice point which is not part of a fermion loop. At every site the worm makes a random decision 
along which link to continue next. These choices are restricted by the condition that the worm is not allowed to hit a fermion loop
or itself (except when it reaches its starting point). For this bookkeeping the worm blocks the sites and links it already contains.
The worm also may end up at a site where the only possible move would be to retrace the previous step. An example is the 
right end of the frozen horizontal dimer inside the top fermion loop in Fig.~\ref{fig1}. 
In such a case the worm is deleted and no update is performed. 
Once the worm reaches its starting point it stops and the worm has identified a closed contour where links with dimers
alternate with empty links. The update is completed by deleting all dimers along the contour and filling all previously empty links
with a dimer. We remark that the dimer worm also includes the simple case of, e.g., rotating a pair of two neighboring
horizontal dimers into a vertical pair. 

Most important, however, is the fact that the worm can also update winding contours of alternating dimers and empty links,
which is not possible with the local pair rotations alone. Only including the dimer worm makes the algorithm ergodic. 
The fact that the dimers give rise to configurations with topological properties can already be seen from the problem 
of filling an empty lattice with dimers: It is known (see, e.g., \cite{dimers}) that in two dimensions the dimer configurations come in 
four different topological sectors which are not connected by local transformations. Thus it is necessary to include  
a global update such as the worm strategy discussed here.

Let us now come to discussing the update of the fermion loops and the plaquette occupation numbers attached to them.
These updates can be done with three types of local steps on single plaquettes which we illustrate in Fig.~3. 
In each of these steps the plaquette occupation number of that plaquette changes as $p(n) \rightarrow p(n) \pm 1$, which 
leads to a changed weight in the dual representation (\ref{zfinal2}). Consequently all the three types of steps 
have to be accepted in a Metropolis step with probability min $\{\rho_\pm, 1\}$ where 
\begin{equation}
	\label{rho_pm}
\rho_+ \; = \; \sqrt{\frac{\eta}{\bar{\eta}}} \ \frac{I_{|p(n_0)+1|} (2 \sqrt{\eta \bar{\eta}})}{I_{|p(n_0)|} (2 \sqrt{\eta \bar{\eta}})} 
\quad , \quad \; \; 
\rho_- \; =  \; \sqrt{\frac{\bar{\eta}}{\eta}} \ \frac{I_{|p(n_0)-1|} (2 \sqrt{\eta \bar{\eta}})}{I_{|p(n_0)|} (2 \sqrt{\eta \bar{\eta}})} \; ,
\end{equation}
and the sign $\pm$ is used according to the change $p(n) \rightarrow p(n) \pm 1$.

\begin{figure}[t!]
\hspace*{-4mm}
\includegraphics[scale=0.35]{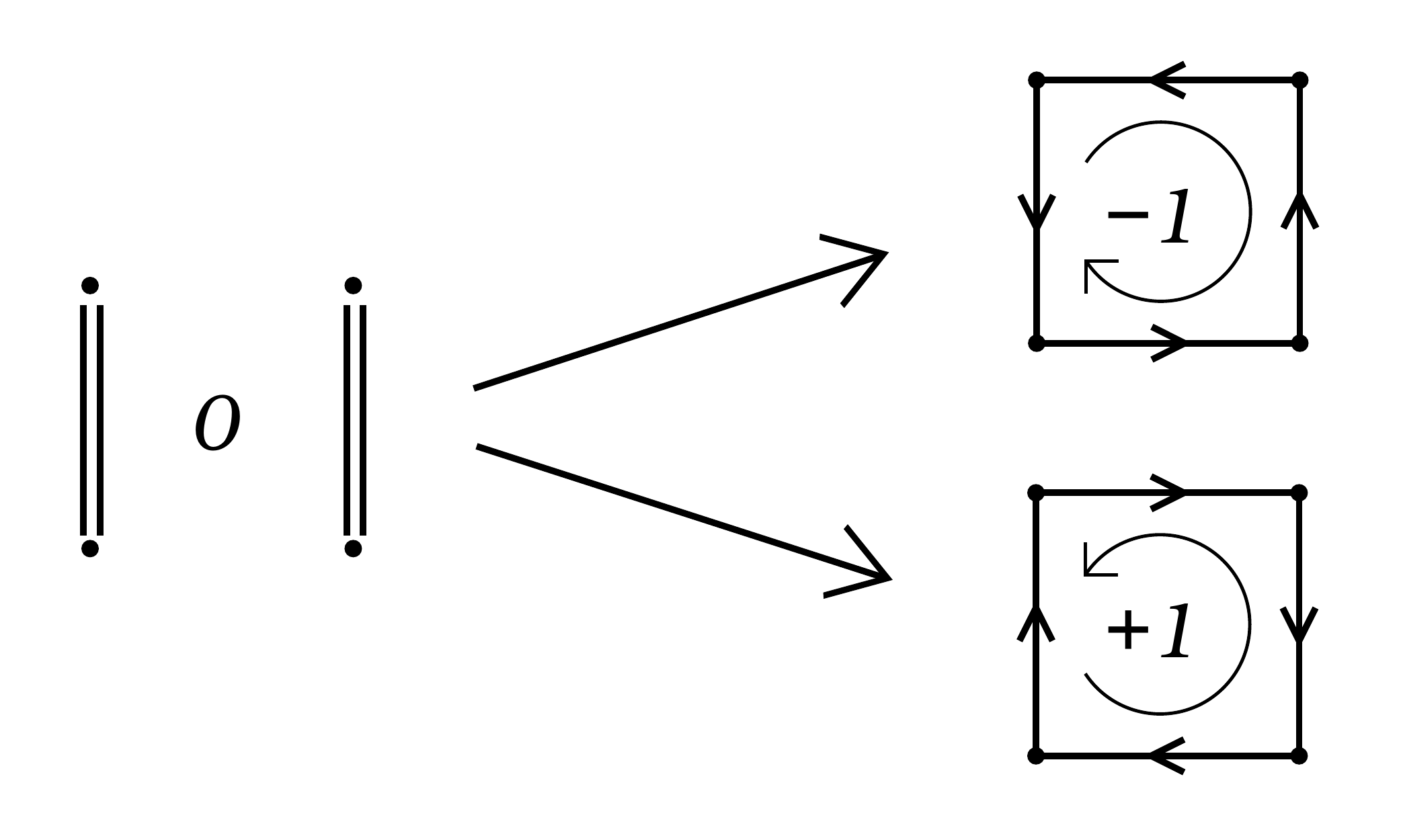}
\hspace{23mm}
\includegraphics[scale=0.35]{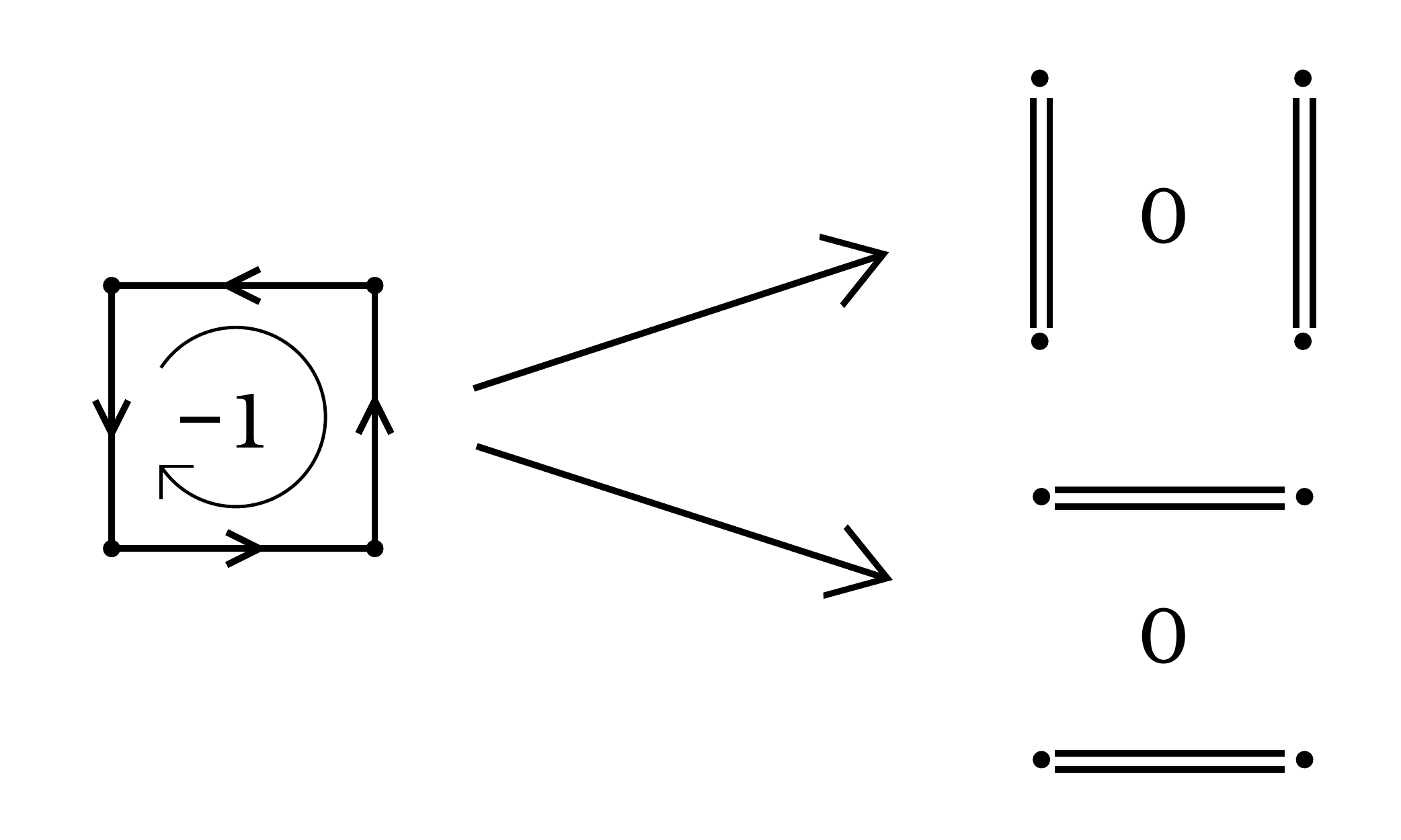}

\noindent
Figure~3.a.: Example of an update step where a pair of dimers is converted into an elementary fermion loop around a plaquette (lhs.\ plot)
and the corresponding inverse move where an elementary fermion loop is converted into a pair of dimers (rhs.). For both, the insertion 
of the fermion loop, as well as for the removal of the fermion loop we have two possibilities which are 
offered with equal a-priori probability. The corresponding plaquette
occupation number $p(n)$ changes to $p(n) \pm 1$. We remark that in all plots of Fig.~3 the changes of the plaquette occupation
number $p(n)$ are always denoted relative to the empty configuration, e.g., $ 0 \rightarrow \pm 1$. In the general case, e.g., 
when the plaquette is inside another loop, the change is $p(n)  \rightarrow p(n) \pm 1$.

\vspace{8mm}

\hspace*{-4mm}
\includegraphics[scale=0.35]{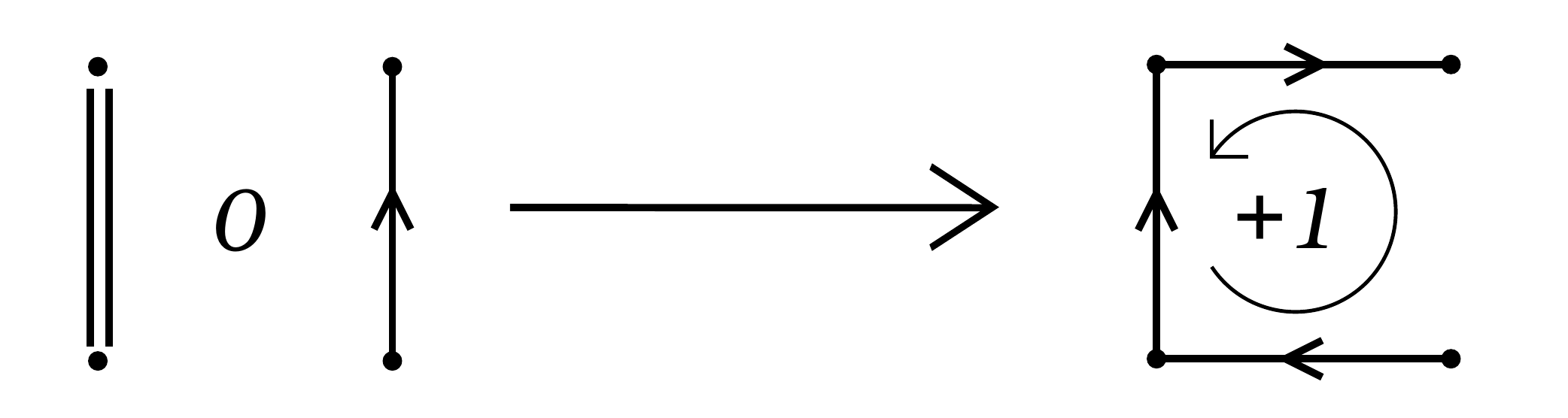}
\hspace{23mm}
\includegraphics[scale=0.35]{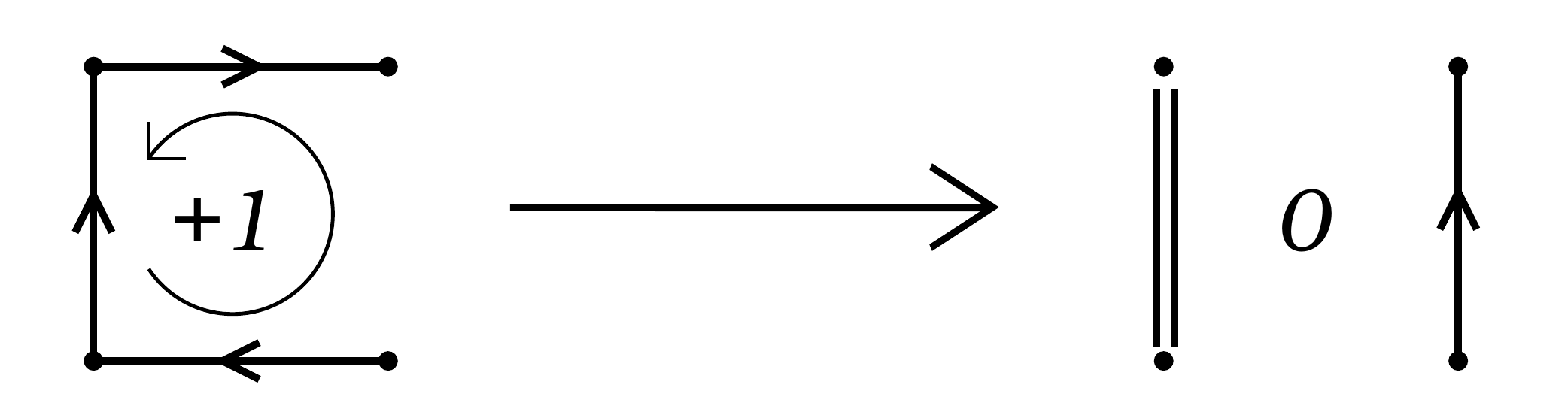}

\noindent
Figure 3.b.: Example of an update step where a fermion loop expands and a dimer is deleted (lhs.) and the corresponding inverse move 
where a fermion loop shrinks and a dimer is placed (rhs.). The corresponding plaquette
occupation number $p(n)$ changes to $p(n) \pm 1$.

\vspace{8mm}

\begin{center}
\includegraphics[scale=0.35]{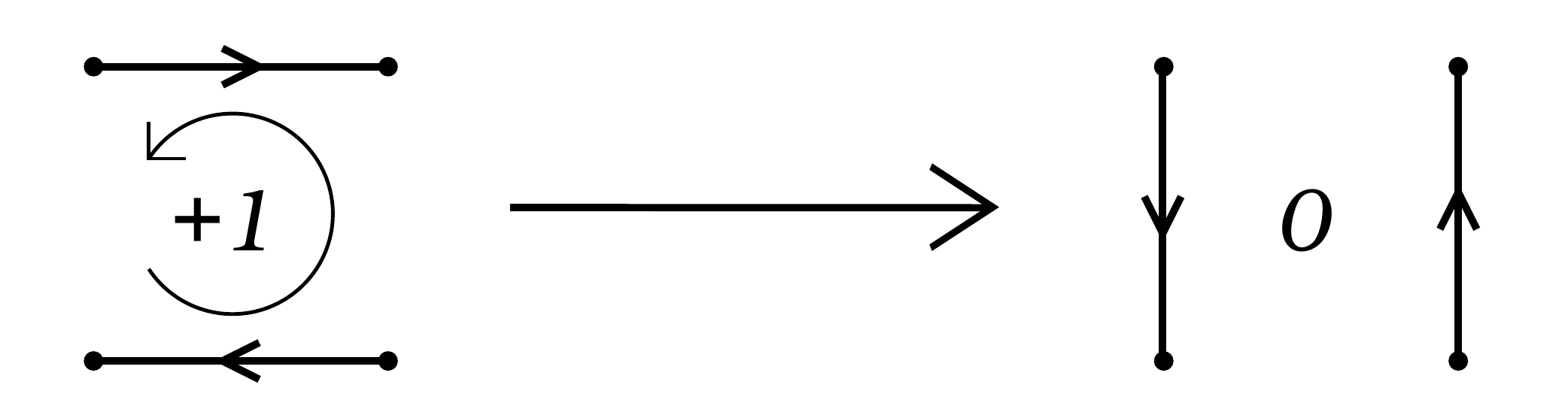}
\end{center}

\noindent
Figure 3.c.:  Example of an update step where we either join two fermion loops or split a single fermion loop into two separate loops (which 
of these two situations applies depends on the global connectivity properties of the fermion loops). The corresponding plaquette
occupation number $p(n)$ changes to $p(n) \pm 1$ in these steps.

\setcounter{figure}{3}
\end{figure}

A sweep of our algorithm runs through all plaquettes and attempts 
a change of the fermion variables on that plaquette together with the corresponding plaquette occupation number $p(n)$. 
Depending on the situation the algorithm finds on the plaquette it works on, the algorithm suggests one of the 
types of moves illustrated in Fig.~3, and it is easy to see that these three types of steps together with the dimer worm described above 
give rise to an ergodic update. 

The first type of step (Fig.~3.a., lhs.\ plot) is for the situation when the plaquette is occupied by 
a pair of parallel dimers. In this case the algorithm offers to replace the two dimers by an elementary fermion loop around the
plaquette, where both possible orientations are offered with equal a-priori probability. The plaquette occupation number changes 
from $p(n)$ to $p(n) \pm 1$ and the step is accepted with the Metropolis probability min$\{\rho_\pm,1\}$ with $\rho_\pm$ given 
in (\ref{rho_pm}). The inverse step is illustrated in the rhs.\ plot of Fig.~3.a.: here the algorithm attempts to remove an elementary
fermion loop around a single plaquette and to replace it with a pair of dimers, where again the horizontal and the vertical possibility
are suggested with equal a-priori probability and accepted with probability min$\{\rho_\pm,1\}$  depending on the orientation of the 
loop that is replaced by the pair of dimers.

The second type of update step for fermion loops and plaquette occupation numbers is illustrated in the lhs.\ plot of 
Fig.~3.b. Here one adds a detour to a loop by removing a dimer that sits on the same plaquette as the segment of 
fermion loop. The plaquette occupation
number is changed, $p(n) \rightarrow p(n) \pm 1$ according to the orientation of the new path of the fermion loop of that plaquette
(in Fig.~3.b.\ we show the situation for an example relative to the empty configuration $p(n) = 0$). The move is again accepted 
with  probability min$\{\rho_\pm,1\}$. The rhs.\ plot of Fig.~3.b.\ shows the inverse process, i.e., a detour around 
a plaquette is taken out of a loop and the fermion constraint is satisfied by placing a dimer.

Finally, in Fig.~3.c.\ we illustrate the third possible case the algorithm attempts to update at a plaquette: Two anti-parallel 
segments of fermion loops on two opposing links of a plaquette. In this case the algorithm offers to delete the two segments
and to insert them on the other two links of the plaquette. This leads to a configuration where two loops have been joined, or a 
loop has been separated into two loops (depending on the connectivity properties of the loop(s) the two original loop segments belong 
to). Again the plaquette occupation number changes as $p(n) \rightarrow p(n) \pm 1$, giving rise to a Metropolis acceptance probability
of min$\{\rho_\pm,1\}$.

In all other cases, e.g., at plaquettes with a piece of loop around only one corner plus the endpoint of a dimer on the opposite corner, or
a plaquette with two pieces of loop running in the same direction on opposite links of the plaquette, the algorithm does not 
propose a change. 

We stress that the moves for changing the fermion loops together with the plaquette occupation numbers we discussed here
cannot introduce a non-vanishing net-winding number of loops. The local moves on the plaquettes can only create pairs of
fermion loops that wind (spatially or temporally), where one loop in the pair winds forward and the other one backward 
(in space or time). Since forward and backward loops appear in pairs, no net winding number is introduced. However, this is 
exactly what the Gauss law enforces: a non-zero winding would correspond to a system that is not charge neutral. It is easy 
to see that the dual formulation implements the Gauss law in a simple geometrical way: a single loop that winds cannot be saturated with 
occupying plaquettes. Having understood that winding loops are not compatible with the constraints, it is trivial to see that the 
three types of steps shown in Fig.~3 give rise to an ergodic update for the fermion loops and the gauge fields in the one flavor case. 
In the case of two flavors of opposite charge discussed in the next section, we can have pairs of winding fermion loops and there 
additional strategies (i.e., worms for double fermion loops) will be needed.  

For the simulation of the system with the topological term it is advantageous to add another update: a global shift of all 
plaquette occupation numbers. In this step we offer the change $p(n) \rightarrow p(n) \pm 1 \; \forall n$, where the sign $\pm$
is randomly chosen with equal probability. Obviously such a gobal shift of the plaquette occupation numbers leaves all 
constraints intact, since the additional flux on neighboring plaquettes cancels along the link they share. The  
step is accepted with Metropolis probability min$\{\rho_{g,\pm},1\}$, where 
\begin{equation}
\rho_{g,\pm} = \left(\sqrt{\frac{\eta}{\bar{\eta}}}\right)^{\pm V} \ \prod_n 
\frac{I_{|p(n)\pm 1|}(2 \sqrt{\eta \bar{\eta}})}{I_{|p(n)|}(2 \sqrt{\eta \bar{\eta}})} \; .
\label{plaqshift}
\end{equation}
We stress that this step is not necessary for ergodicity, since such a global shift can also be generated by a loop growing to the size
of the whole lattice, such that it wraps around the periodic boundaries and closes back onto itself. However, this is a rare event, and the 
explicit global change of the plaquette occupation number speeds up the propagation in configuration space.

\subsection{Results and checks}

Having developed the dual representation and a corresponding new update strategy it is paramount to test these. Here we use two 
types of tests. For the case of $\theta = 0$, i.e., when no topological term is coupled, we can use a standard simulation, since the
case of $\theta = 0$ is free of the complex action problem also in the conventional formulation. 
In the standard simulation the fermions are integrated out and the
corresponding fermion determinant is used as an additional weight factor (together with $e^{-S_G[U]}$) for the Monte Carlo sampling
of the gauge configurations $U$. The fermion determinant is real and positive, since the eigenvalues of the staggered Dirac operator
come in complex conjugate pairs. For the 2-dimensional case of the Schwinger model the discretized Dirac operator is a relatively
small matrix, such that for generating the reference data we could simply compute the fermion determinant with a standard library. 

The second test we implemented was an exact summation over all possible fermion configurations on a small lattice, i.e., fillings 
of the lattice with dimers and fermion loops, and a numerical sum over all plaquette occupation numbers compatible with a 
given configuration of the fermion loops. The details for obtaining this exact result on small volumes are described in the appendix. The
exact summation is possible also for $\theta \neq 0$ and allows us to test the dual simulation also in the case where the
complex action problem is present.

Before we come to presenting the tests and the numerical results, let us briefly discuss the parameters used in our 
simulations. We approach the continuum limit as in \cite{scalarqed2} by simultaneously sending to infinity 
the inverse gauge coupling $\beta$ and the lattice volume $N_S N_T$ at a fixed ratio 
\begin{equation}
R \; = \;  \frac{\beta}{N_S \, N_T} \; = \; \, \mbox{const} \, \; .
\end{equation}
For the tests shown here a typical value of the constant would be $R = 0.1$ and we considered lattice sizes up to $N_S = N_T = 24$.
The values of $\theta$ were chosen in a range between $-3\pi$ and $+3\pi$. Note that the geometrical definition of
the topological charge gives an integer charge $Q[U]$ only in the continuum limit, such that also the $2\pi$-periodicity
in $\theta$ emerges only in that limit and it is useful to monitor a larger interval of $\theta$-values than just the principal branch 
of $[-\pi, \pi]$ (actually since observables are either even or odd in $\theta$, already the interval $[0,\pi]$ is sufficient 
in the continuum limit). The dual simulation of the one flavor model typically uses between $10^6$ and 
$10^7$ measurements 
which are decorrelated by a mix of 20 sweeps of local updates for loops and plaquettes, 10 dimer worms and 10 global updates for 
the gauge plaquettes. At each point in parameter space we use $5 \times 10^4$ cycles of  these mixed 
updates for equilibration. The errors we show are statistical errors determined with single elimination jackknife.

\begin{figure}[t!]
\hspace*{-4mm}
\begin{center}
\includegraphics[scale=0.35]{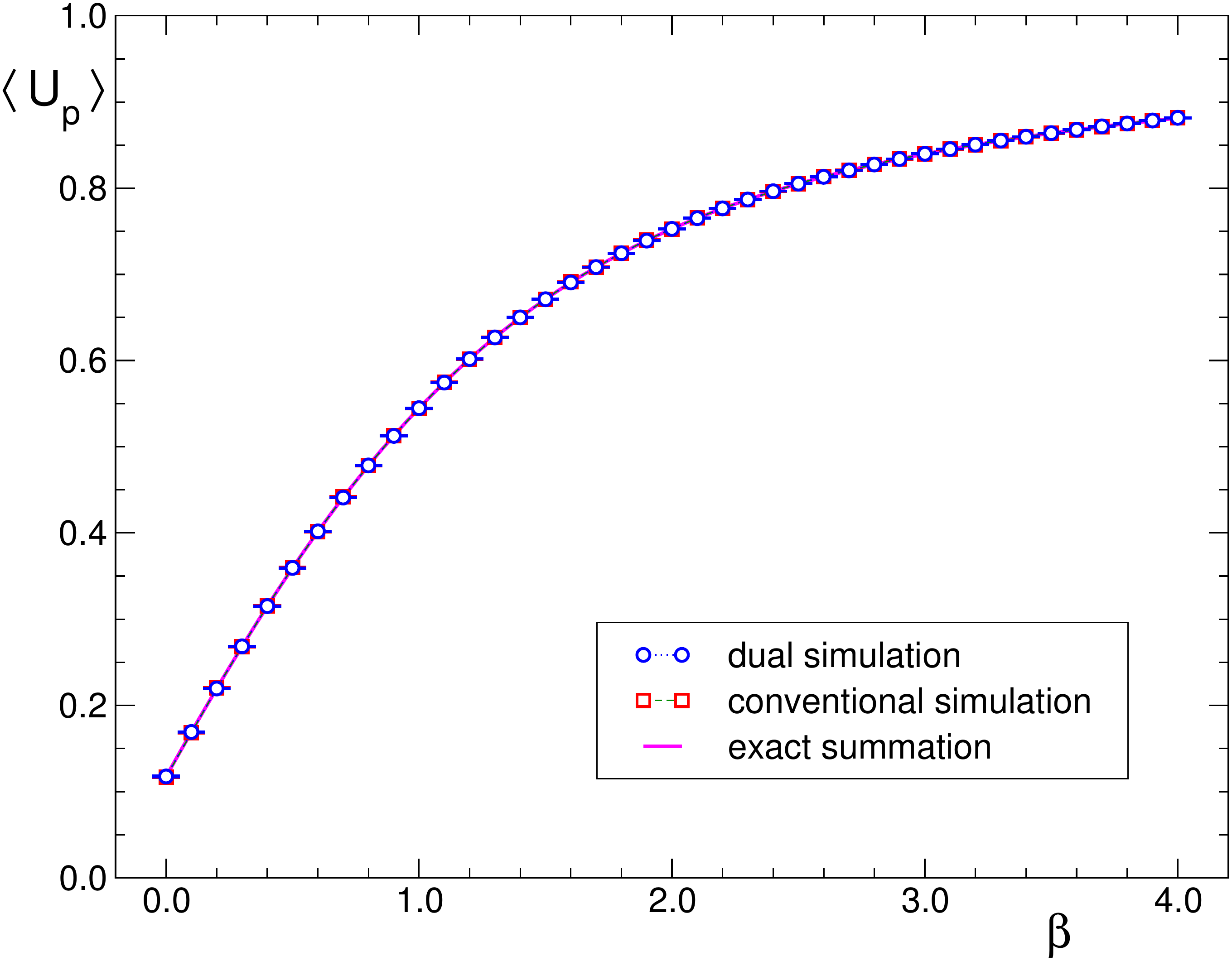}
\end{center}
\caption{The plaquette expectation value in the one flavor model at $\theta = 0$ 
as a function of $\beta$. We compare the results from the dual simulation (circles) 
with the data from a conventional simulation (squares) and the results from an exact summation (full curve) 
on a $4 \times 4$ lattice. We find perfect agreement 
of the three results thus confirming the correctness of the dual representation and the update strategy.}
\label{check_1F_1}
\end{figure}

We begin the discussion of our tests with a comparison of all three approaches possible at $\theta = 0$, i.e., conventional simulation,
exact summation and dual simulation. In Fig.~\ref{check_1F_1} we show the results of the three calculations as a function of $\beta$ for 
lattice size $4 \times 4$ (larger lattices are too big for the exact summation). It is obvious that the dual simulation (circles), the conventional 
simulation (squares) and the results of the exact summation (full curve) agree perfectly, 
supporting the correctness of the dual representation at $\theta = 0$ and the numerical implementation of the dual simulation.

\begin{figure}[t!]
\hspace*{-4mm}
\begin{center}
\includegraphics[scale=0.5]{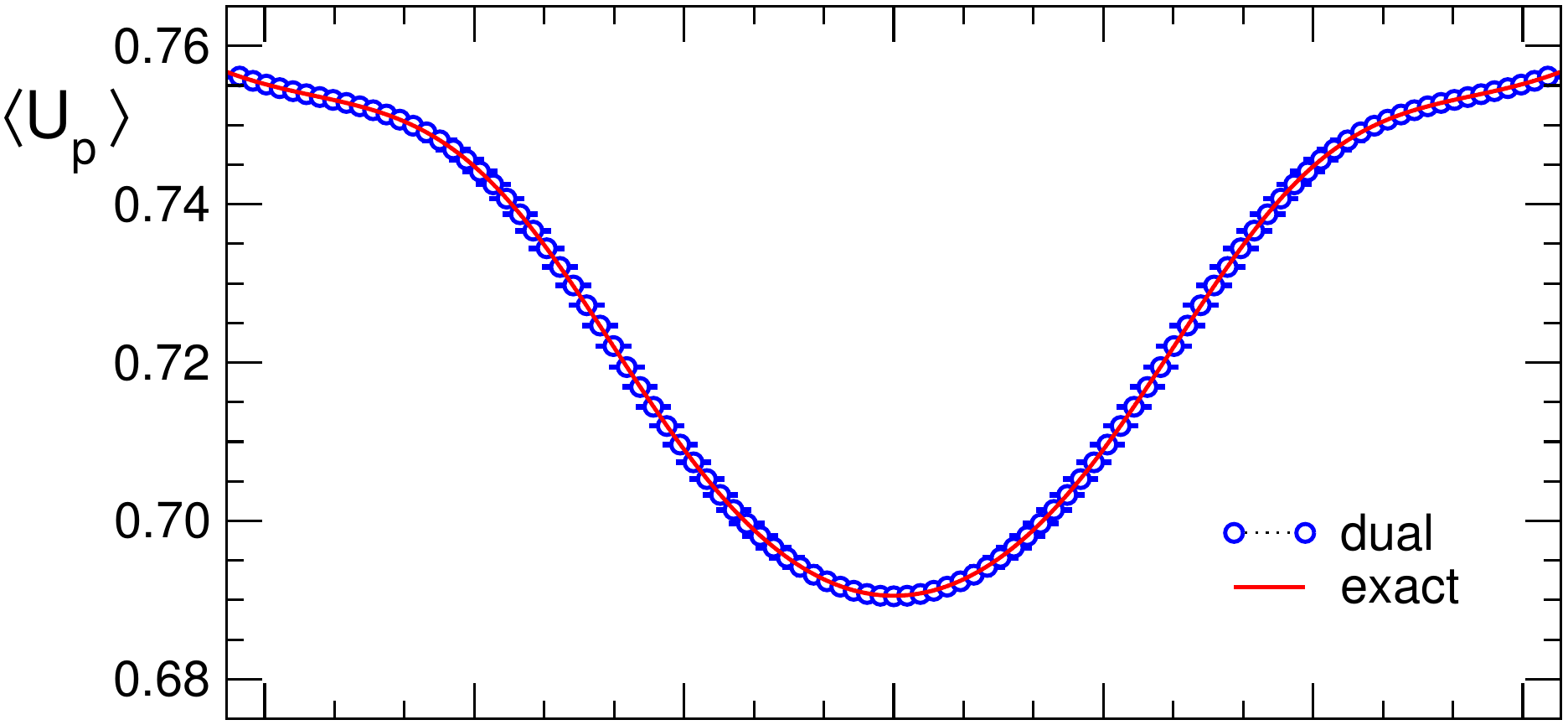}
\hspace*{-0.7mm}\includegraphics[scale=0.5]{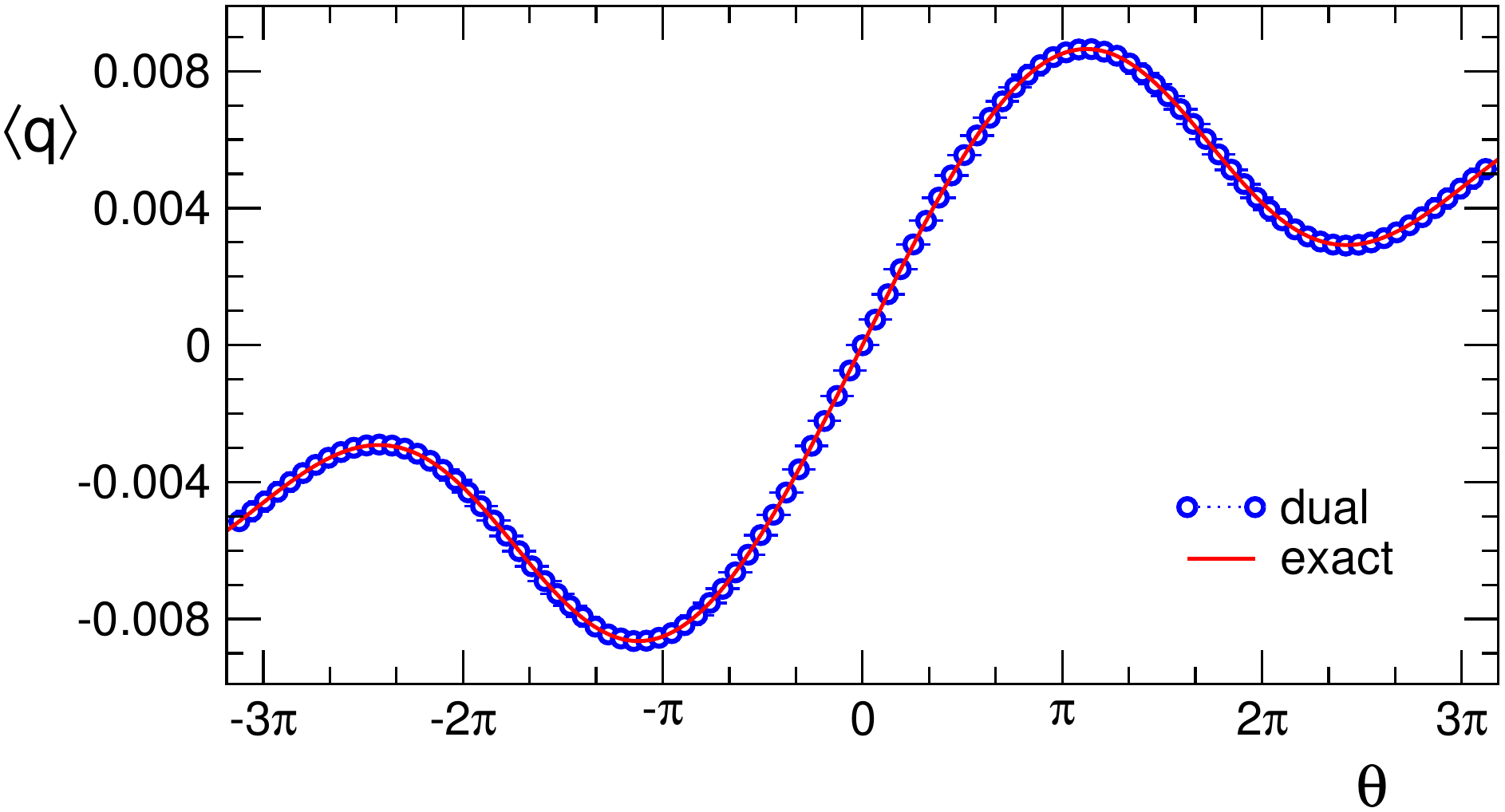}
\end{center}
\caption{Plaquette expectation value and topological charge density in the one flavor model as a function of
$\theta$ at fixed $\beta = 1.6$ and lattice size $4 \times 4$. We compare the data of the dual simulation (circles) 
with the results of the exact summation (full curve) 
on a $4 \times 4$ lattice. We find perfect agreement 
of the results thus confirming the correctness of the dual representation and the update.}
\label{check_1F_2}
\end{figure}

\begin{figure}[t!]
\hspace*{-4mm}
\begin{center}
\includegraphics[scale=0.53]{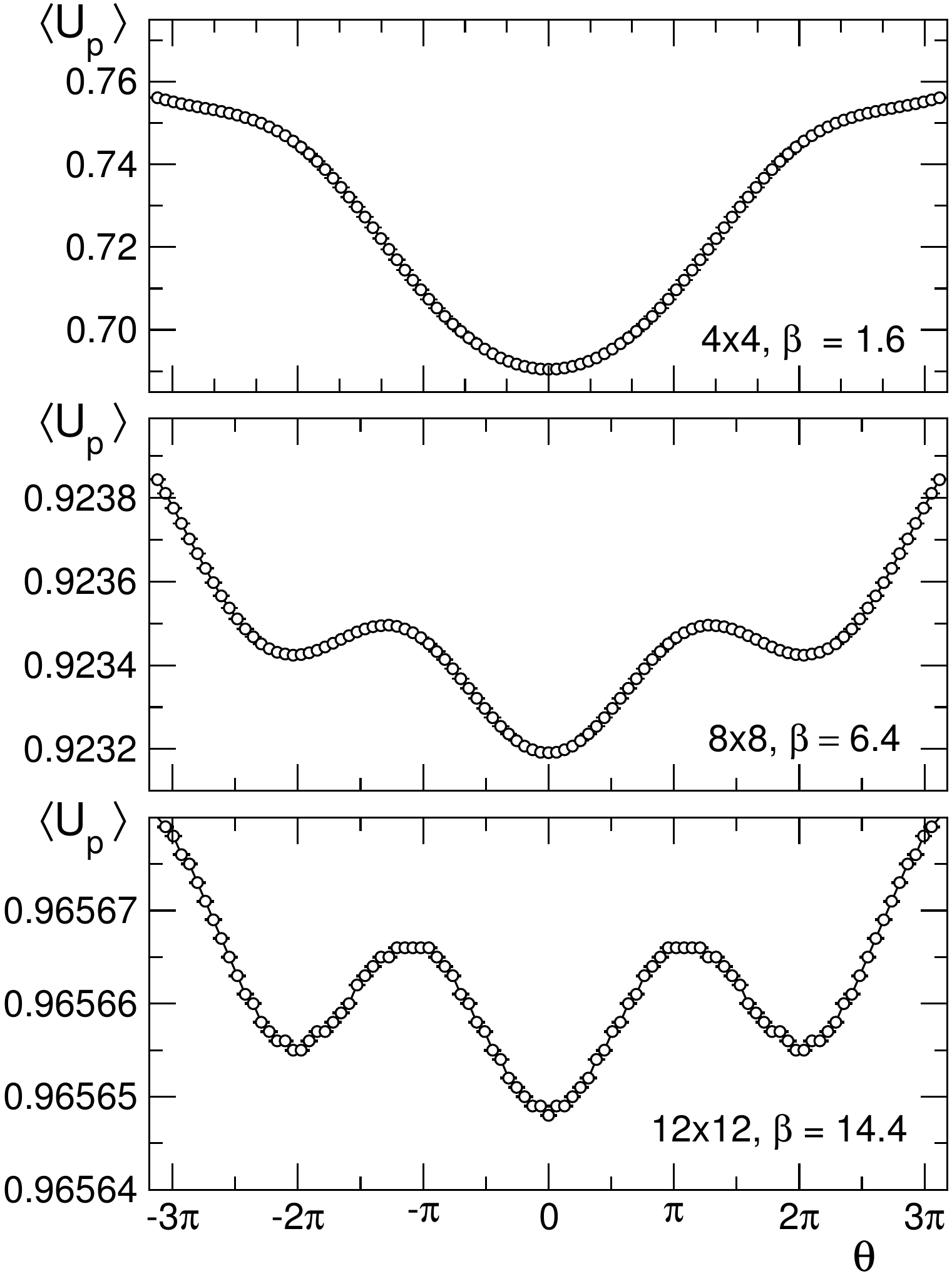}
\hspace*{1mm}\includegraphics[scale=0.53]{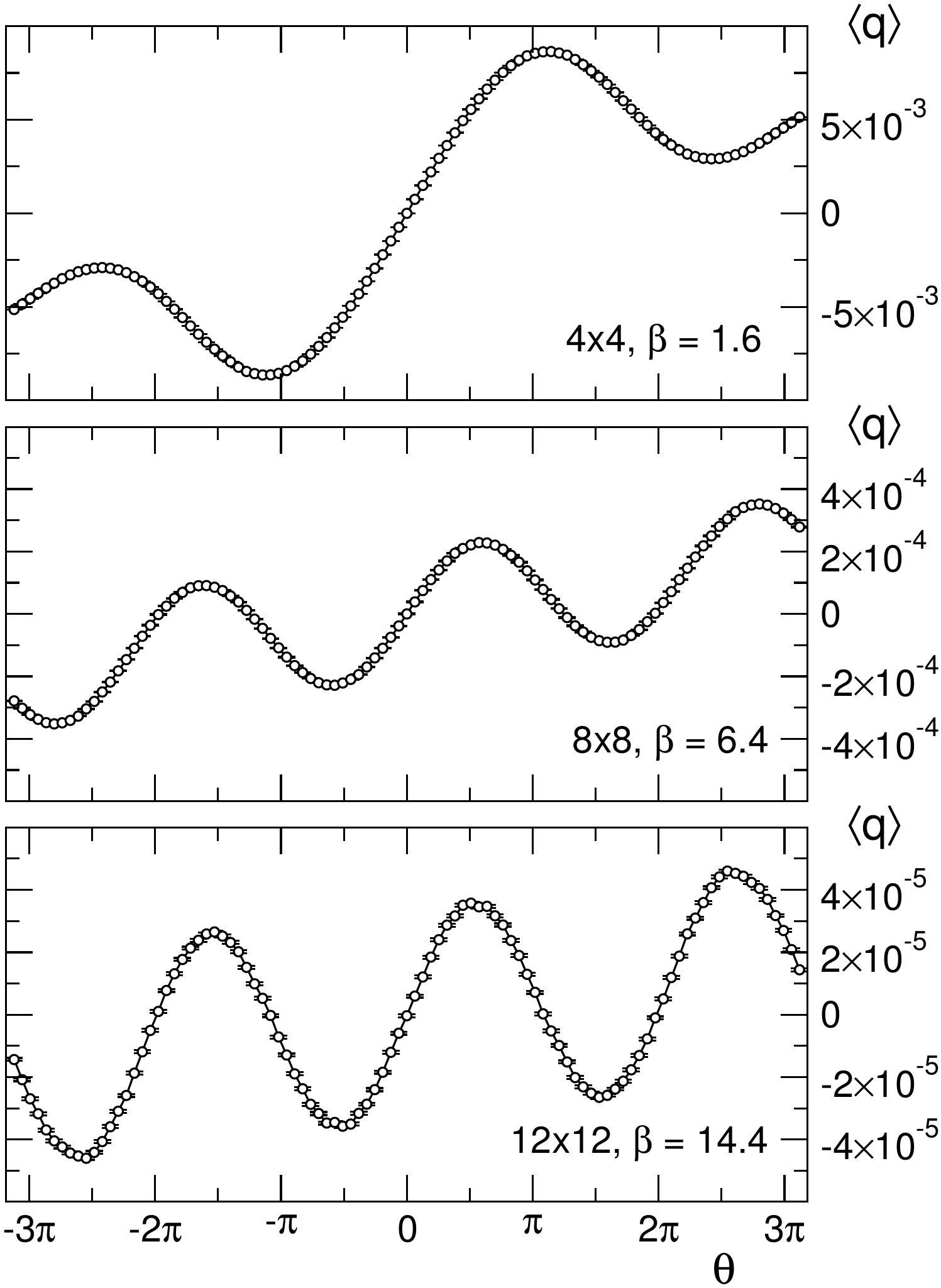}
\end{center}
\caption{Plaquette expectation value (lhs.\ plot) and topological charge density (rhs.) towards the continuum limit. We show the 
quantities as a function of $\theta$ for three different values $\beta = 1.6, 6.4$ and 14.4 (top to bottom) and increase the volume 
such that $R = \beta / N_S N_T$ remains constant at $R = 0.1$. 
We observe the emergence of the $2\pi$-periodicity expected in the continuum limit.}
\label{2pi_periodicity}
\end{figure}

For non-zero vacuum angle $\theta$ we can no longer use the conventional approach, but tested our dual simulation 
program against the results of the exact summation on a small lattice at finite $\theta$ the results of this test are presented 
in Fig.~\ref{check_1F_2} where we show the plaquette expectation value $\langle U_p \rangle$ (top plot) and the topological 
charge density $\langle q \rangle$ as a function of $\theta$. The data from the dual simulation (circles) and the results
of the exact summation (full curve) agree perfectly, confirming our update strategy for the dual representation also at non-zero $\theta$. 

We remark at this point that the fact that the comparison to the exact summation is possible only for small volume is not 
necessarily a disadvantage: on a small lattice configurations that wind in some way around the periodic boundary conditions  
play a larger role and if not taken into account correctly by the update algorithm would lead to stronger deviations between 
the Monte Carlo data and the results from exact summation. In our case these configurations are fermion loops that wind around 
the compactified directions (in the one flavor case these can only be pairs of forward and backward winding loops), as well as 
sheets of plaquette occupation numbers that cover the entire torus (simply changing all plaquette occupation numbers 
from $p(n)$ to $p(n) \pm 1$ in a admissible configuration creates such a sheet which again gives rise to an admissible configuration). 
Furthermore, as already mentioned, also dimer configurations are known \cite{dimers} 
to come in topologically distinct sectors that necessarily imply a 
global strategy such as our dimer worm for an ergodic update.
The fact that we find perfect agreement on the $4 \times 4$ lattice indicates that also configurations winding around the periodic 
boundaries are taken into account correctly by the fermion/gauge algorithm and the worm for the dimers.

Having checked the correctness of the algorithm and the dual representation let us briefly analyze the physics of the vacuum angle
that emerges from the formulation of the model with massless staggered fermions and the field theoretical definition of the topological 
charge. The study of the $\theta$-dependence in scalar QED$_2$ \cite{scalarqed2} suggests that also here a $2\pi$-periodicity 
of the dependence of observables on the vacuum angle $\theta$ should emerge only in the continuum limit.
This is indeed manifest in Fig.~\ref{2pi_periodicity} where we show the plaquette expectation value $\langle U_p \rangle$ (lhs.\ plots) 
and the topological charge density $\langle q \rangle$ (rhs.) as a function of $\theta$. From top to bottom we approach the continuum 
limit by increasing $\beta$ and the volume. It is obvious that both observables start to develop the expected $2\pi$-periodicity, 
as can, e.g., be seen by observing that the distance between two maxima (or minima) approaches the value $2\pi$.

However, there is an important difference between scalar QED$_2$ and the massless Schwinger model. In the latter case one can use a 
chiral rotation and the anomaly of the fermion determinant to show that if one of the fermion masses vanishes (note that for staggered 
fermions we have to compare to the two-flavor continuum results), the physics of the continuum model is independent of 
$\theta$. Thus it is interesting to analyze whether, and in which form this independence of $\theta$ emerges when approaching the 
continuum limit with the formulation with massless staggered fermions and the field theoretical definition 
of the topological charge\footnote{For studies of the $\theta$ 
dependence for the massive lattice Schwinger model see, e.g., \cite{onogi}.}. 
When inspecting the results for $\langle U_p \rangle$ (lhs.\ plots in Fig.~\ref{2pi_periodicity}) one finds that they resemble a parabola
overlaid with oscillations, while the results for $\langle q \rangle$ resemble an inclined straight line overlaid with oscillations. 
We first note that in both cases the non-oscillating part becomes more flat towards the continuum limit: 
for $\langle U_p \rangle$ the curvature of the parabola
decreases when increasing $\beta$ and the same holds for the slope of the data for $\langle q \rangle$. Furthermore one clearly sees 
that also the amplitudes of the oscillations decrease quickly toward the continuum limit: the amplitude of the oscillation of 
$\langle U_p \rangle$ (lhs.\ plots) between the minimum at $\theta = 0$ and the first maximum roughly has the values 
$6 \times 10^{-2}$,  $3 \times 10^{-4}$ and $2 \times 10^{-5}$ for $\beta = 1.6, 6.4$ and 14.4, and the amplitude of the oscillation of 
$\langle q \rangle$ (rhs.\ plots) between the two extrema near $\theta = 0$ roughly decreases as $1.7 \times 10^{-2}$,  
$4.4 \times 10^{-4}$ and $7 \times 10^{-5}$. This qualitative analysis illustrates how the $\theta$ dependence disappears towards 
the continuum limit, as expected for the continuum model with one or more massless fermions. This supports the results
about the correct implementation of the anomaly by the lattice formulation with massles staggered fermions \cite{durr} 
(although other observables might be sensitive to the order of continuum- and chiral limit).

\begin{figure}[t!]
\begin{minipage}{0.5\textwidth}
\vspace*{4mm}
\includegraphics[scale=0.55]{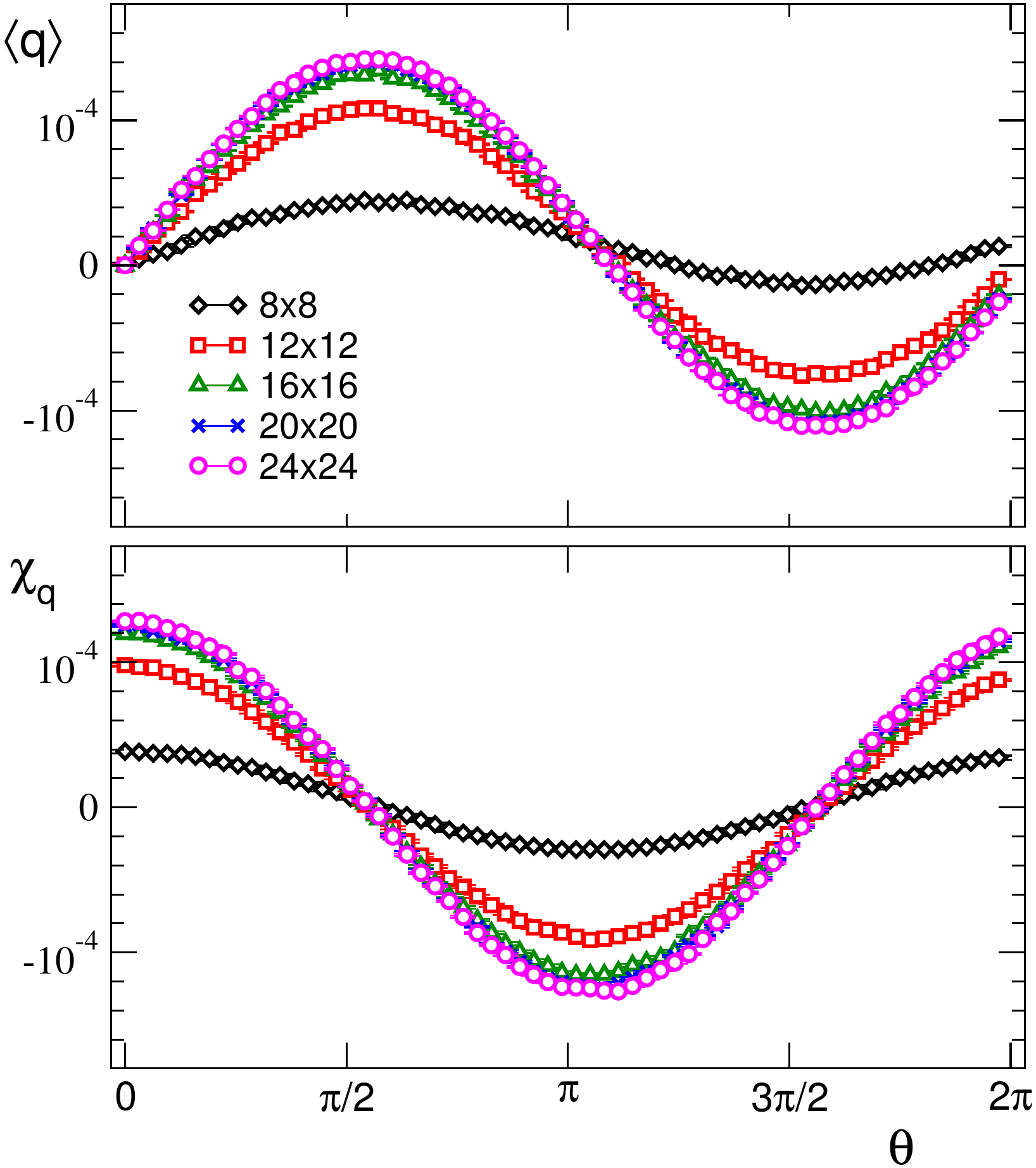}
\end{minipage}
\hspace*{20mm}
\begin{minipage}{0.45\textwidth}
\includegraphics[scale=0.45]{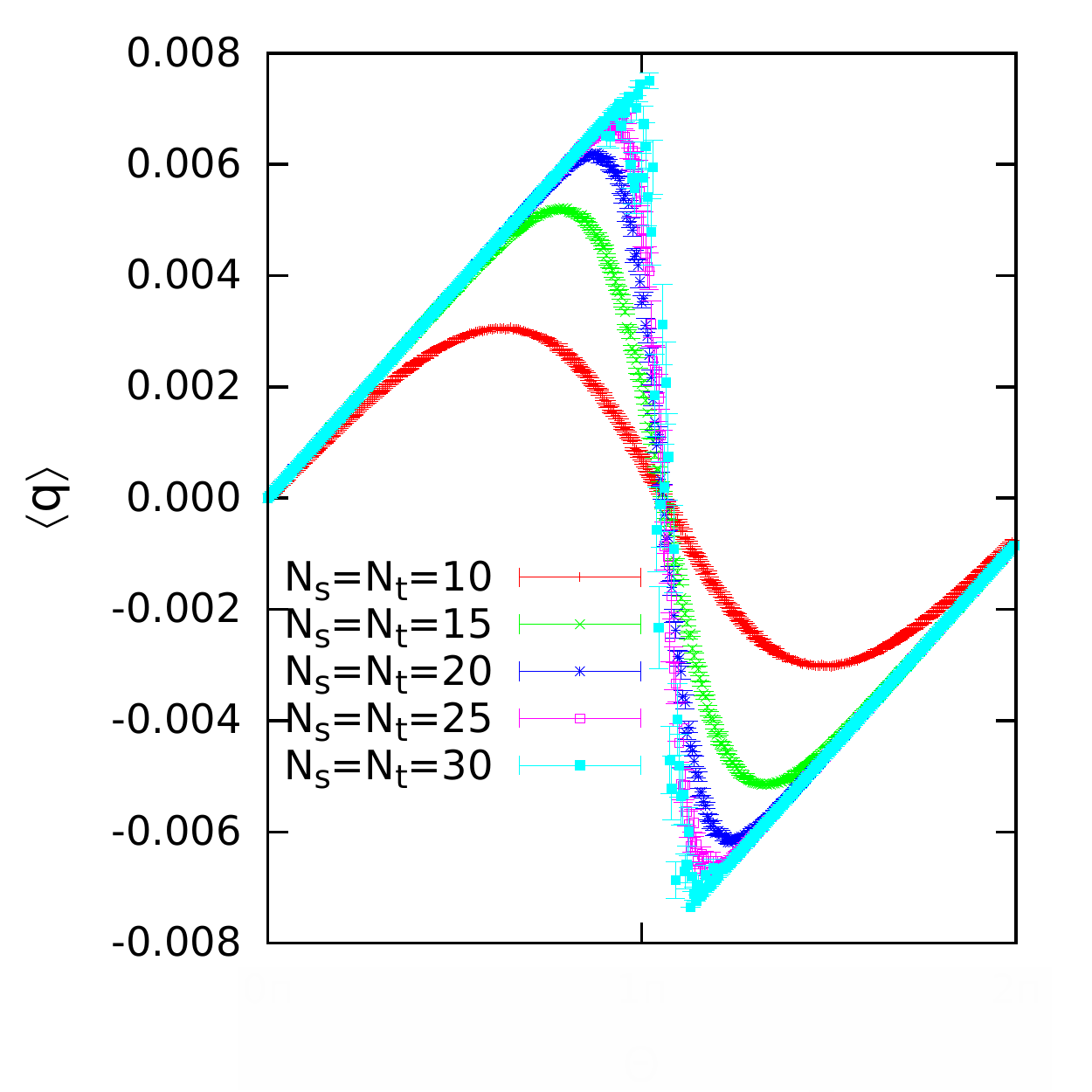}
\vskip-8mm
\hspace*{0.5mm}
\includegraphics[scale=0.436]{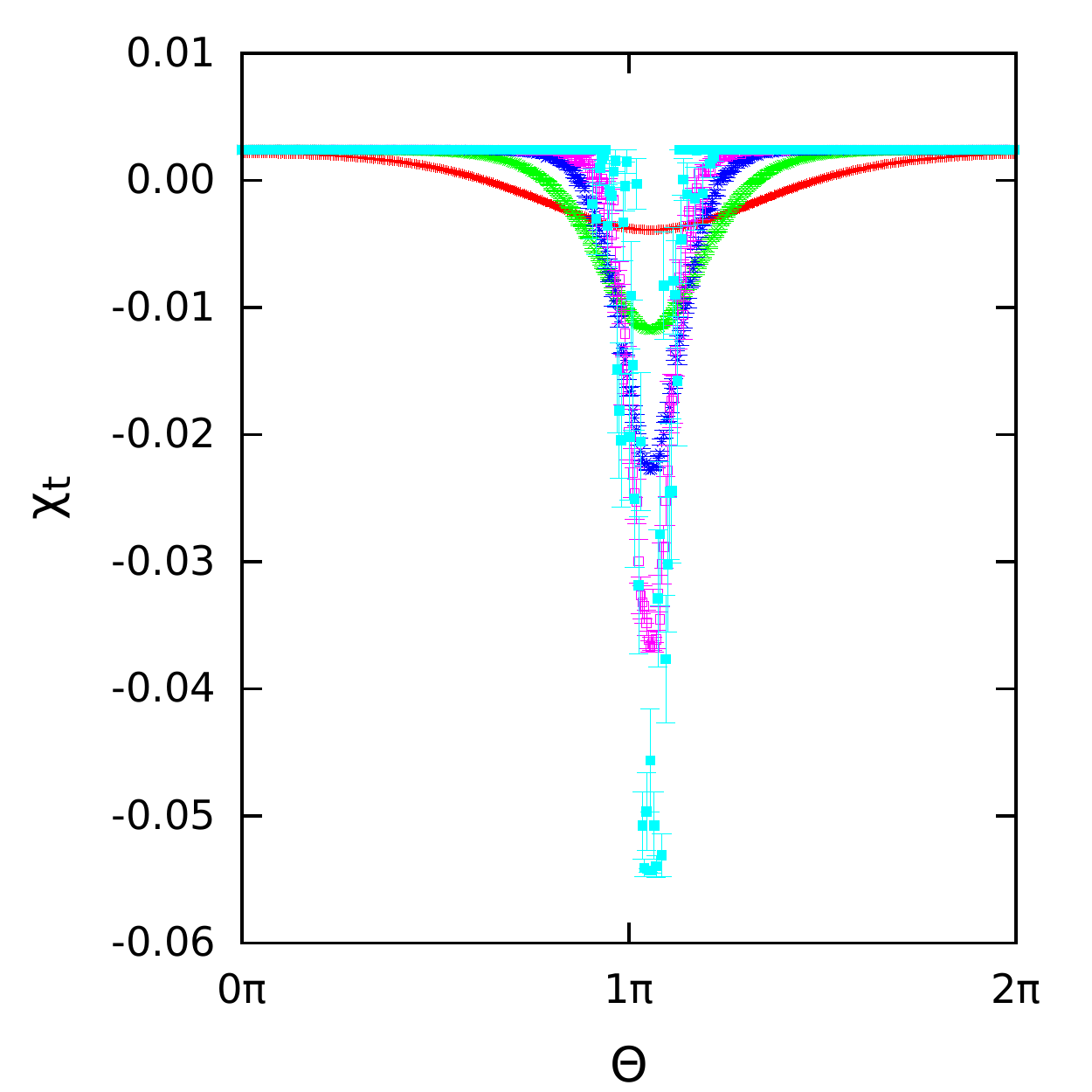}
\end{minipage}
\caption{Topological charge density $\langle q \rangle$ (lhs., top plot) and the corresponding susceptibility $\chi_q$ (lhs., bottom)
as a function of the vacuum angle $\theta$. At fixed $\beta = 10.0$ we compare the results for
different lattice sizes to study the volume dependence. For comparison the rhs.\ plots show the same quantities for scalar 
QED$_2$, where indeed a first order behavior emerges at $\theta = \pi$ (plots from \cite{scalarqed2}).}
\label{transition_pi}
\end{figure}

Finally we address a related question in our exploratory study, namely the possibility of a phase transition at 
$\theta = \pi$ when approaching the thermodynamical limit at fixed inverse coupling $\beta$. For pure U(1) gauge theory in 
2 dimensions, as well as for scalar QED$_2$ a first order phase transition emerges
at $\theta = \pi$, which can, e.g., be seen in a jump in $\langle q \rangle$ and a diverging topological susceptibility. 
These findings were established also with the geometrical definition of the topological charge (see, e.g., \cite{scalarqed2} 
and the two plots on the rhs.\ of Fig.~\ref{transition_pi}): a first order jump emergences in $\langle q \rangle$
at $\theta = \pi$, and the maximum in the topological susceptibility grows proportional to the volume.

In the lhs.\ plots of Fig.~\ref{transition_pi} we show our results for the topological charge density $\langle q \rangle$ (top plot) and
the topological susceptibility $\chi_t$ (bottom) as a function of $\theta$ for the Schwinger model. We work at a 
fixed $\beta = 10.0$ and increase the volume from $8 \times 8$ up to  $24 \times 24$. It is obvious that both observables 
saturate as the volume increases and do not develop a phase transition, different from the results for the same observables 
in scalar QED$_2$ \cite{scalarqed2}. Also this result for the absence of a transition supports the correct approach 
of a $\theta$-independent continuum limit.

\section{Update strategies and results for the two flavor model with chemical potential}
\label{twoparticle}

Having established the validity of our algorithmic approach in the somewhat simpler one flavor case with topological charge, and
having discussed some of the observables, we now come to the two flavor case. Here the focus is on finite
density, i.e., finite chemical potential, and thus we now set the vacuum angle to $\theta = 0$ (although its inclusion in the two flavor
model is straightforward). The algorithmic challenge in the two flavor case is to efficiently update the winding fermion loops that 
carry the dependence on the chemical potentials $\mu_\psi$ and $\mu_\chi$.

\subsection{Steps of the update}

The update of the two flavor model builds on the steps already developed for the one flavor model, since the set of configurations in the 
dual two flavor partition sum (\ref{z2flavordual}) contains products of admissible fermion 
configurations of the one flavor partition sum (\ref{zfinal2}) for each 
of the two flavors. Thus as the first part of our algorithm for both flavors we reuse the sweeps of the single flavor updates as discussed 
in the previous section. We alternately perform the local loop deformations illustrated in Fig.~3 and the dimer worms for each of the flavors,
combined with offering a global shift of all plaquette occupation numbers. These updates access 
all configurations that have zero net winding 
number of fermion loops around both, spatial as well as temporal direction for each of the flavors. 

However, as the example in Fig.~2 shows, in the two flavor case also configurations are 
admissible where the net winding number around space 
and time has the same non-zero value for fermion loops of both flavors. 
In the example of Fig.~2 this is the temporal (= vertical) fermion double loop on the right hand side of 
the configuration. We will now present a worm algorithm that is capable of inserting and removing such 
fermion double loops. We stress at this point, that the fermion double loop in Fig.~2 is a very special configuration 
with non-zero net winding number, but it is easy to see that the local loop 
deformations of the previous section generate all possible instances of configurations in the same winding class. In particular,
the two strands of the double loop can become split when, e.g., the update steps shown in Fig.~3.b.\ are applied to one of the two flavors.
 
The main challenge of inserting a winding loop is to find a path where the loop can be placed such that the fermion constraints are 
obeyed. To achieve such an insertion we first run a worm that identifies a closed contour where dimers of both colors 
alternate\footnote{A second possibility would be to find a loop where we have dimers of both colors on the same links alternating 
with empty links, but since the single flavor dimer loops which we run switch between these two cases, the second possibility needs
not be included explicitly.}. Once such a loop is identified, we replace it by a fermion double loop, where both possible orientations are 
chosen with equal probability. If the fermion double loop has zero temporal winding the change is always accepted. Otherwise 
we accept it with probability min $\{ 1, e^{ W(\mu_\psi + \mu_\chi)N_T\!} \}$, where $W$ is the temporal winding number of the double loop. 
This step is illustrated in the lhs.\  plot of Fig.~8, together with the inverse step, where a fermion double loop (again identified 
with a worm) is replaced by a closed contour of dimers (rhs.). Together with the single flavor updates of Section 3 this 
already constitutes an ergodic algorithm for the two flavor model. 

\begin{figure}[t!]
\vspace*{-6mm}
\begin{center}
\hspace*{-2mm}
\includegraphics[scale=0.39]{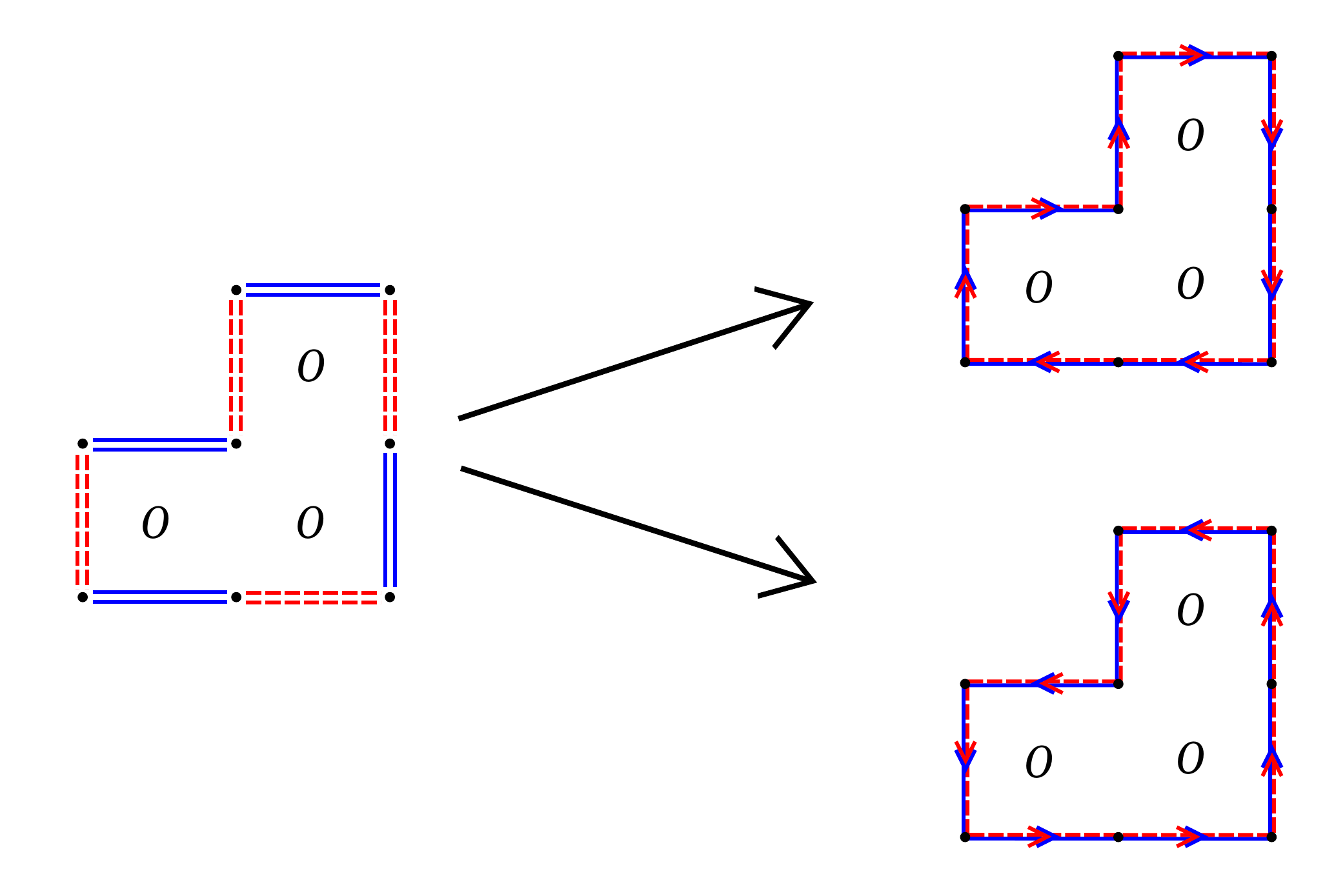}
\includegraphics[scale=0.39]{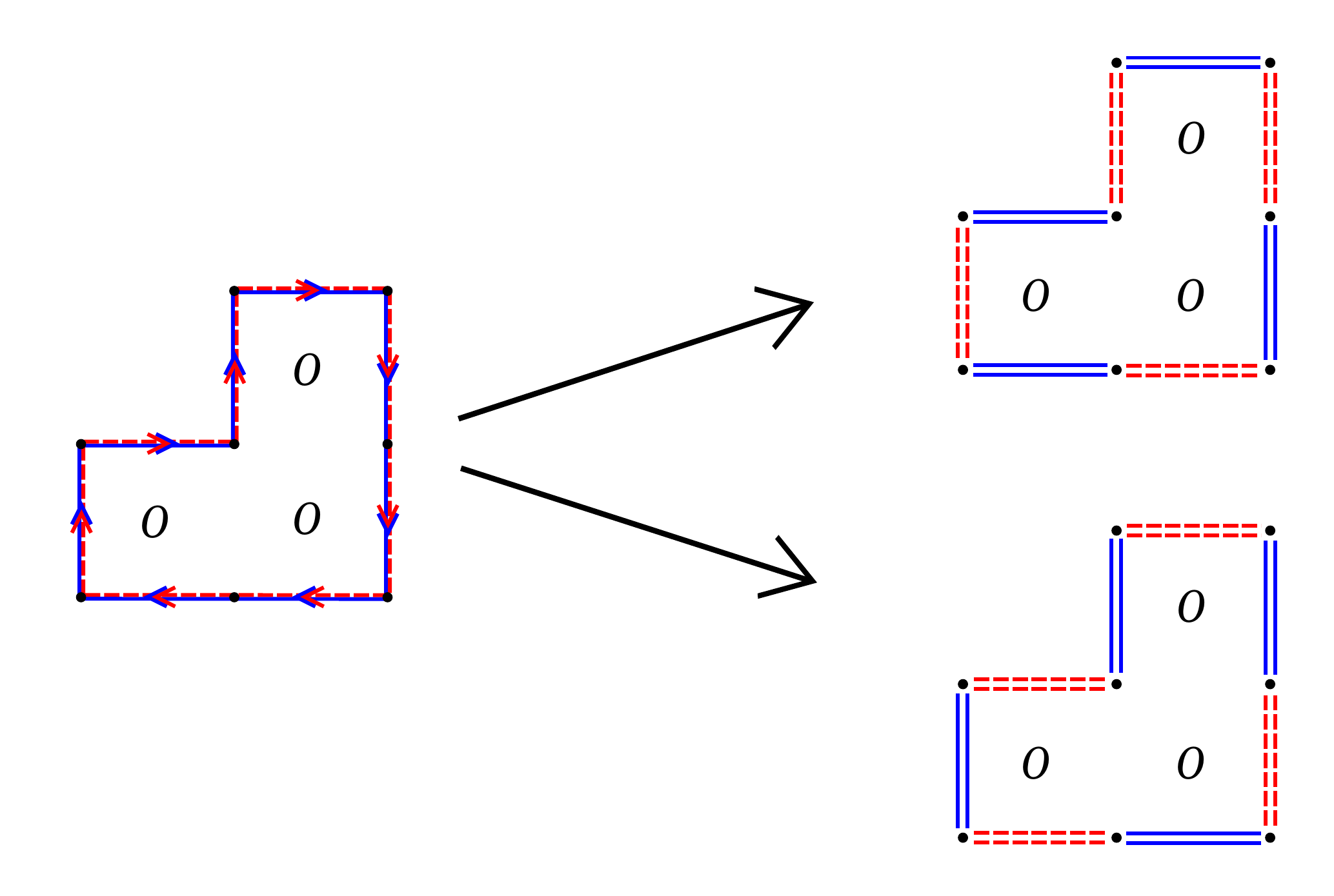}
\end{center}
\vspace{-3mm}
\caption{Worm algorithm for fermion double loops. The worm identifies a closed chain where dimers of both flavors alternate and 
replaces the chain by a fermion loop (lhs.\ plot). The inverse step is a worm that identifies a closed double fermion loop and replaces
it by a chain of alternating dimers of both flavors.}

\vskip8mm

\begin{center}
\begin{center}
\includegraphics[scale=0.31]{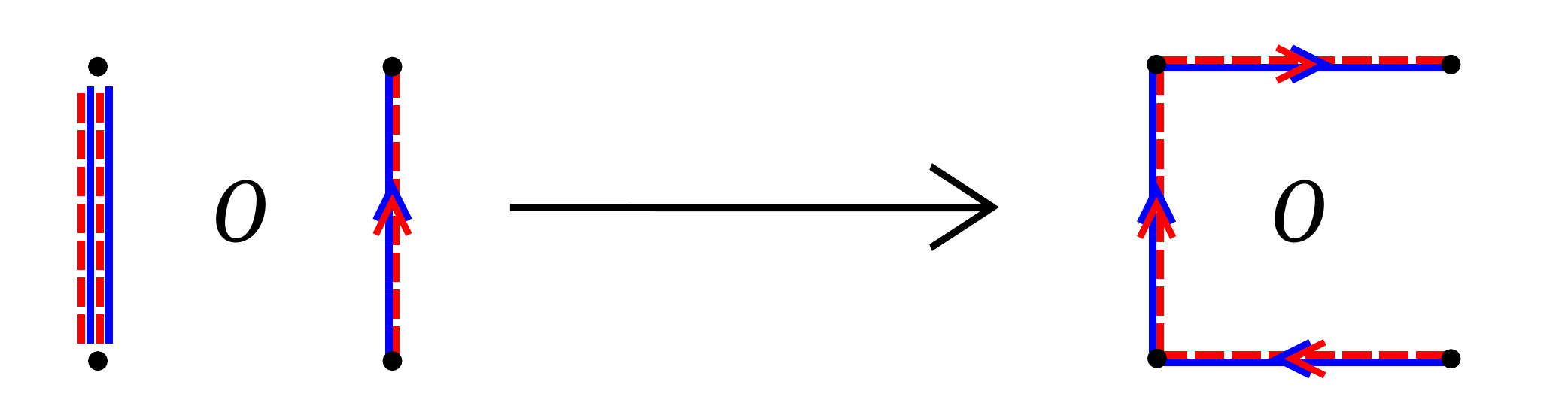}
\hspace{28mm}
\includegraphics[scale=0.31]{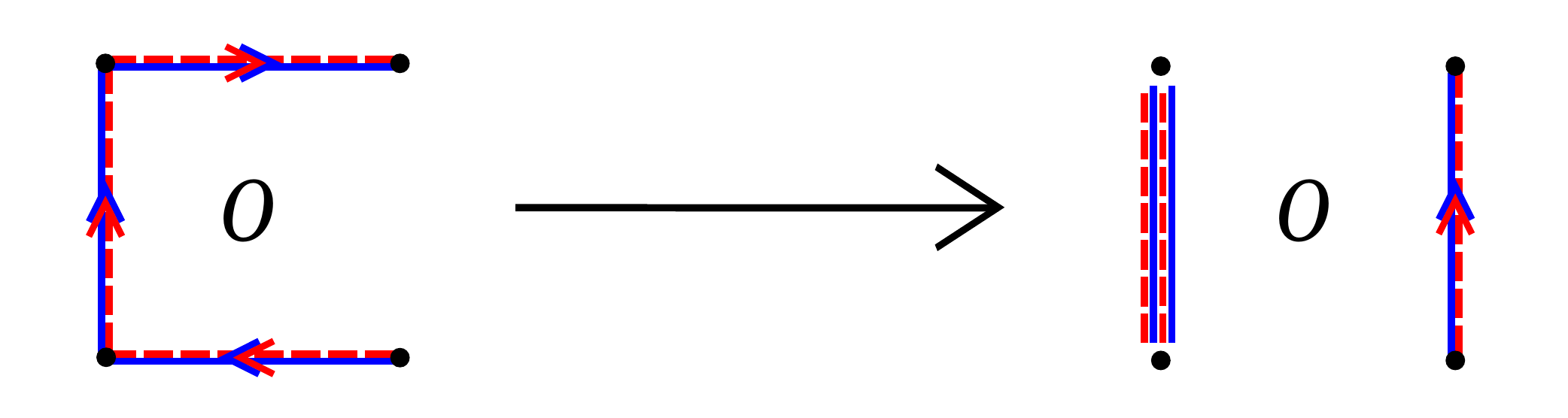}
\end{center}

\vskip2mm

\noindent
Figure~9.a.: Inserting a detour in a double fermion loop after 
removing a dimer pair (lhs.\ plot) and the corresponding inverse step (rhs.).

\vskip8mm

\begin{center}
\includegraphics[scale=0.31]{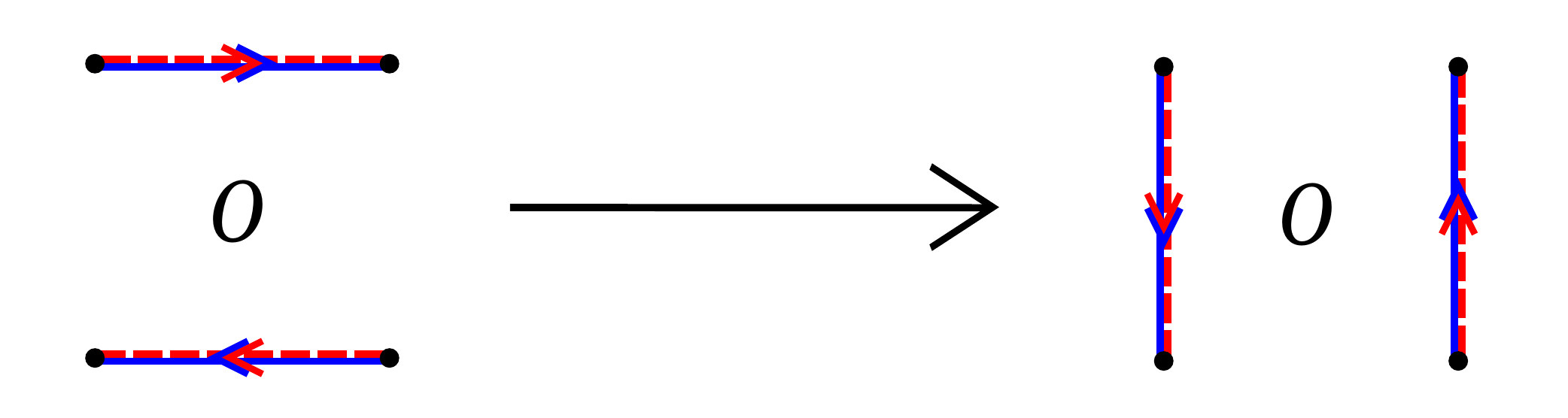}
\end{center}

\noindent
Figure~9.b.: Joining (splitting) of a double fermion loop.

\end{center}

\setcounter{figure}{9}
\end{figure}

However, one may increase the efficiency of the update by also including steps that locally deform double fermion loops (see 
the steps in Fig.~9). In principle these steps lead to changes that are possible also via two other mechanisms: A complete removal 
of the double loop and a subsequent reinsertion after a dimer change, or as a sequence of two (or more) single flavor steps with a
change of plaquette occupation numbers. The former possibility often (depending on the loop) has a low probability and the second
one becomes strongly suppressed at small $\beta$. Thus we include steps for the fermion double loops that are similar to the 
local steps in the single flavor update, but do not require a worm search or the activation of plaquettes in an intermediate step. In 
Fig.~9.a.\ we show the step where a double dimer is replaced by a detour of the fermion double loop (lhs.) and the corresponding
inverse step (rhs.). Fig.~9.b.\ shows a step where we join two double loops or split them (depending on the overall connectivity
properties of the double fermion loop). 

\begin{figure}[t!]
\hspace*{0mm}
\hspace*{-2mm}
\includegraphics[scale=0.4]{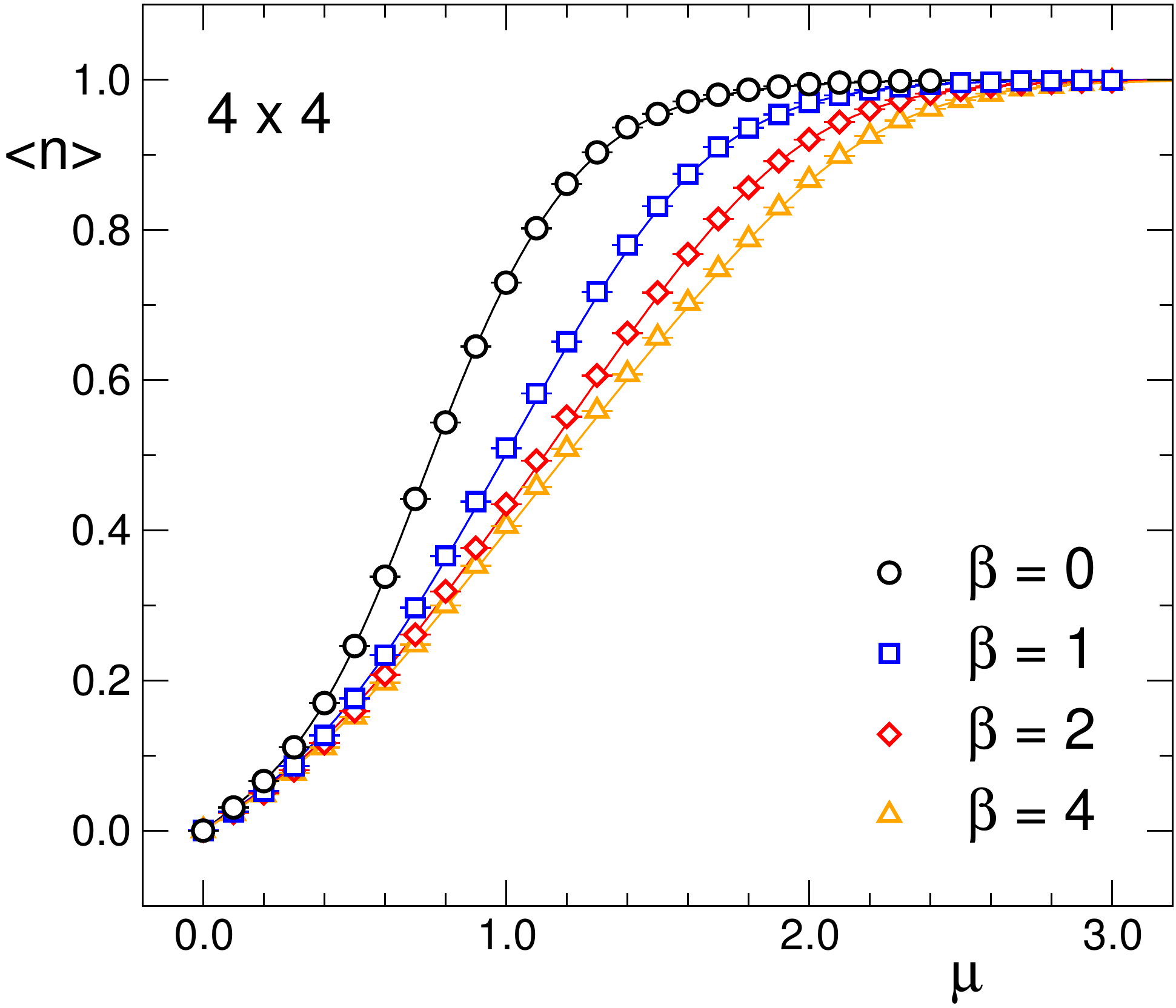}
\hspace{3mm}
\includegraphics[scale=0.4]{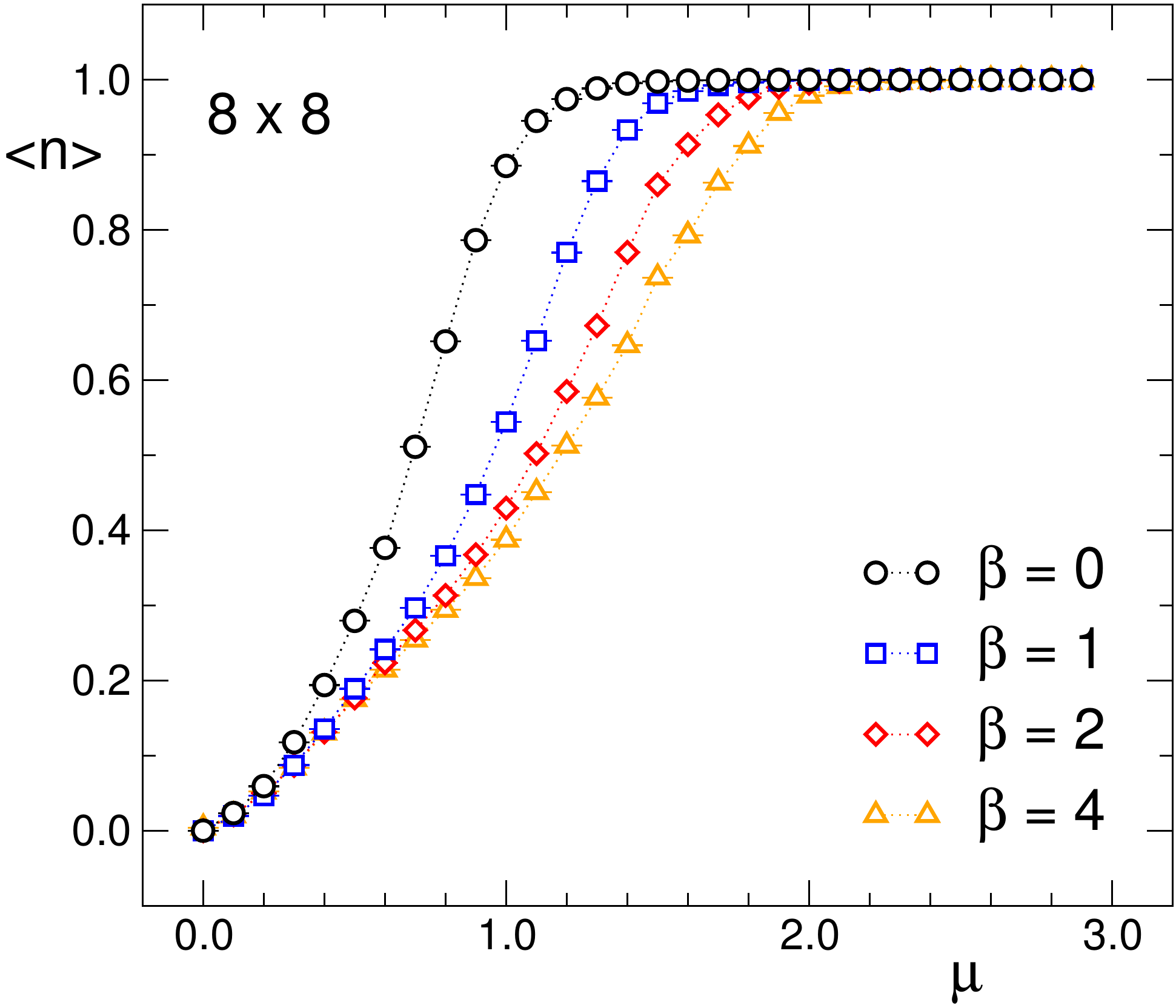}
\caption{The particle number density $\langle n \rangle$ as a function of the chemical potential, for different values of $\beta$. 
For lattice volume $4 \times 4$ (lhs.\ plot) we compare the data from the dual simulation (symbols) with the results of the 
exact summation (full curves) and find excellent agreement. For the $8 \times 8$ lattice (rhs.) we connect the data from the 
simulation with dashed lines to guide the eye.}
\label{n_vs_mu}
\end{figure}

\subsection{Results and checks}

For the actual simulation of the two flavor model we use the following combined sweeps: One combined update sweep consists of
one sweep of the local updates for each of the two flavors, as well as worm update sweeps for dimers of both flavors as
discussed in Section 3.1. This is combined with one global plaquette shift (compare Eq.~(\ref{plaqshift})), one sweep of local neutral loop 
updates as described in Fig.~9, as well as one winding loop update sweep as shown in Fig.~8. For each parameter set
we equilibrate the system by performing $10^5$ to $10^6$ of these combined sweeps and for the evaluation of observables we use 
between $10^6$ and $4 \times 10^7$ configurations separated by 6 to 20 combined sweeps for decorrelation. 

As for the case of the one flavor model with topological term we begin the discussion of our simulation results with a comparison 
to the results of an exact summation of the partition sum on a small lattice (see the appendix for details of the summation). 
In the lhs.\ plot of Fig.~\ref{n_vs_mu} we show the particle number density $\langle n \rangle$ as a function of the chemical potential 
$\mu_{\psi} = \mu_{\chi} \equiv \mu$ for a $4 \times 4$ lattice. The symbols represent the results of our dual simulation for 
$\beta = 0, 1, 2$ and 4, and the full curves are the corresponding reference data from the exact summation. Obviously, for all 
values of $\beta$ the dual simulation results agree very well with the curves from the exact summation, which indicates that the 
dualization and the dual simulation were implemented correctly. The curves for $\langle n \rangle$ saturate at 1 as expected for fermions.

In the rhs.\ plot of Fig.~\ref{n_vs_mu} we show our results for the particle number density $\langle n \rangle$ as a function of 
$\mu$ for lattice size $8 \times 8$, again comparing the results for $\beta = 0, 1, 2$ and 4. Here we do not have reference data 
from an exact summation and we simply connect the data points that represent the results from the dual simulation to guide the eye.
Comparison of the two figures shows that, as expected, the curves for $\langle n \rangle$ become steeper near the inflection point 
when increasing the volume. The limit of infinite lattice size corresponds to the limit of infinite spatial volume at zero temperature, 
where one expects that $\langle n \rangle$ behaves like a step function and our numerical results show that trend. 

\begin{figure}[t!]
\hspace*{-4mm}
\begin{center}
\hspace*{-0.7mm}\includegraphics[scale=0.47]{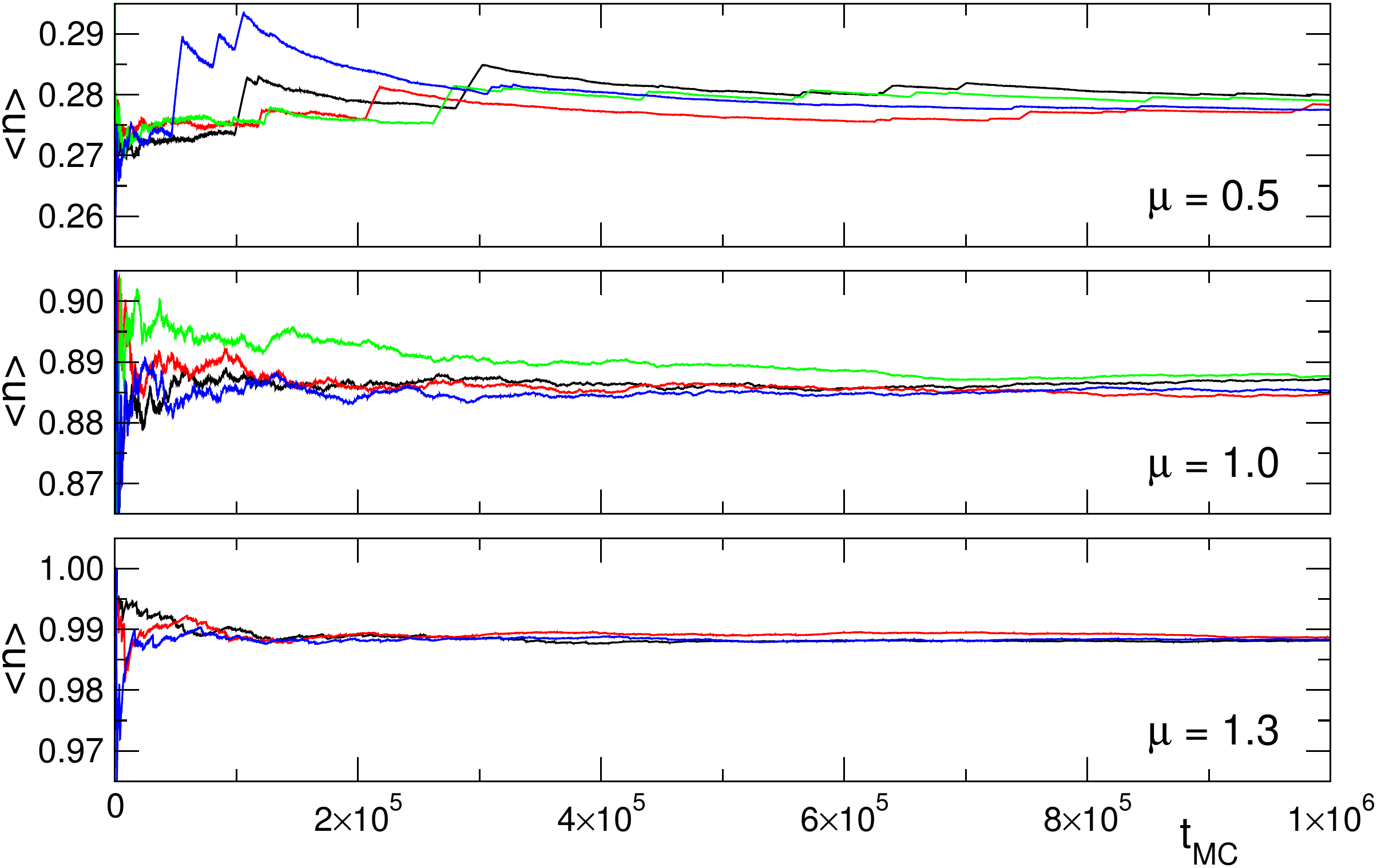}
\end{center}
\caption{Time series for the particle number density $\langle n \rangle$ on a $8 \times 8$ lattice at $\beta = 0$ (i.e., the particle 
number density $n$ is plotted as a function of Monte Carlo time $t_{MC}$ in units of 10 combined sweeps). We compare three 
different values of the chemical potential $\mu = 0.5, 1.0$ and 1.3 (top to bottom) and for each of them 
show the time series for different runs. In particular for small $\mu$ we observe oscillations with low frequency corresponding
to long autocorrelation times. These are due to topologically stabilized double fermion loops (see the discussion in the text).}
\label{timeseries}
\end{figure}

Having established the correctness of the dual simulation and thus the fact that the complex action problem is overcome in principle
by the simulation in terms of worldlines/worldsheets, we also need to address efficiency issues of the approach. During the 
explorative studies presented here we partly encountered very long autocorrelations in our simulations, in particular for $\beta >2$ 
at medium to large values of the chemical potential ($\mu > 0.5$). The reason is the topological nature of the particle number
in the dual representation. Increasing the chemical potential increases the particle number and thus the net winding number of the 
fermionic loops around the compact time direction. At sufficiently high $\mu$ this can lead to loops that wind several times around 
time and then close around the spatial direction. Note that the net winding number has to be the same for both flavors, and in such 
a configuration the winding loops can already occupy a sizable fraction of the lattice, and for large $\mu$ when $\langle n \rangle$ 
approaches 1 each site of the lattice belongs to a winding loop. Clearly such configurations are stabilzed by topology and despite the 
fact that we use a worm strategy for updating the double loops we found that it is very hard to break up such high-winding configurations. 

This topologically stabilized autocorrelation is clearly visible in the time series of the particle number density which we show in 
Fig.~\ref{timeseries}. The particle number density $\langle n \rangle$ is plotted as a function of the Monte Carlo time $t_{MC}$ 
measured in units of 10 combined sweeps. The three plots of Fig.~\ref{timeseries} are for  $\mu = 0.5, 1.0$ and 1.3 (top to bottom) and
in each plot we show the time series for several different runs. The long correlations are obvious, in particular for the smaller two values of 
$\mu$. The plots illustrate that although the complex action problem is solved by mapping the system to a representation with only real 
and positive contributions, the fermionic nature of the system together with the topological representation of the particle number in terms
of the winding number make the Monte Carlo simulation of the system very stiff. In dimensions $D \geq 3$ 
this problem is expected to be much milder, since then the 1-dimensional winding loops do not separate the lattice into domains that
cannot be intersected with another loop.

We close this section with remarking that the dual formulation also allows for simulations of the canonical ensemble.
Having seen that the system is highly constrained and that changing the particle number is very hard in some parameter regions,
canonical simulations, i.e., simulations at fixed particle number, which in the dual representation corresponds to fixed temporal
winding number, might be a powerful alternative approach. In order to implement a fixed winding number $W$ one starts 
the simulation with an initial configuration that has this winding number, e.g., by placing $W$ double fermion loops that wind once. 
In the subsequent update steps one implements a rejection step for the double fermion line worm whenever it tries to cross the 
last time-slice of the lattice, such that the worm cannot change the total winding number. Together with the local steps this updates 
all configurations with fixed winding number $W$ and thus gives rise to a simulation of the canonical ensemble with particle
number $W$ for both flavors. First tests in a bosonic model \cite{inprep} show that in some situations a dual canonical simulation
clearly outperforms a grand canonical simulation and canonical simulations in the dual formulation might solve the problem with the 
long autocorrelations. 

\section{Summary and concluding remarks}

In this paper we present the first dual simulation of the massless Schwinger model at finite vacuum angle $\theta$ 
and non-zero chemical potential. 
In the dual representation the partition sum has only real and positive contributions, such that the complex action problem is solved.
However, in terms of the new variables, the system is highly constrained. In particular the new fermionic variables, 
the dimers and fermion loops have to obey the Pauli principle, which in the dual formulation requires that each site of the lattice is 
either run through by a loop or is the endpoint of a dimer. These constraints require new update strategies and in this paper we explore
and evaluate different steps that combine into a suitable algorithm. We remark that in the massive case one also has monomers
that can be used for filling the lattice, and since they saturate the fermionic constraints on a single site alone the constraints for the 
fermions are less rigid in the massive case (however, negative sign contributions re-appear \cite{schwinger_dual}).

More specifically we here study the one flavor model with a topological term (i.e., $\theta > 0$) at zero density and the two flavor model 
with finite chemical potential at $\theta = 0$. The update for the one flavor model combines local loop deformations, a worm strategy 
for the update of the dimers and a global shift of the plaquette occupation numbers. The results of the dual simulation are cross checked 
against a conventional simulation at $\theta = 0$, and against an exact summation on small lattices. We verify that the algorithm is 
ergodic and reproduces the reference data with high precision, thus establishing the correctness of the approach. In a first small physics 
application of the new method we analyze how the chosen formulation implements the $\theta$-independence of physics in the continuum
limit and analyze the difference to scalar QED$_2$ for the behavior at $\theta = \pi/2$.

Generalizing the update strategy of the one flavor model, we construct an ergodic algorithm for the two flavor model, by adding a worm
update for the winding double loops that couple to the chemical potential, as well as additional local steps for deforming double loops. 
Again we compare the results from the dual simulation with the outcome of an exact summation on small lattices and confirm ergodicity
and correctness of the dual approach in the two flavor case. When exploring the parameter space at finite $\mu$ we found that for
some parameter sets the system becomes quite stiff due to topological stabilization: increasing the chemical potential enforces high
winding numbers for the fermion loops such that a large fraction of lattice sites is run through by loops that wind around the compact 
time (and space) directions. Obviously this gives rise to configurations that are hard to change - even with worm strategies. Improving 
the sampling of loop configurations in that parameter regime is a challenging task that goes beyond the exploratory first study 
presented here. A possible future strategy for overcoming the stiffness problem of the 
grand canonical approach at large $\mu$ are canonical worldline techniques, which we briefly discuss.

\vskip5mm
\noindent
{\bf Acknowledgments:} 
We thank Stephan D\"urr, Thomas Kloiber, Vasily Sazonov and Tin Sulejmanpasic for many interesting discussions.
This work is partly supported by the Austrian Science Fund FWF Grants.\ Nr.\ I 1452-N27 and I 2886-N27, as well as 
DFG TR55, {\sl ''Hadron Properties from Lattice QCD''}. Daniel G\"oschl is supported by the FWF DK 
W1203 {\sl ''Hadrons in Vacuum, Nuclei and Stars''} and Alexander Lehmann was supported by the Institut f\"ur Theoretische 
Physik of the Humboldt-Universit\"at zu Berlin.

\section*{Appendix: Exact summation on a small lattice}
\begin{figure}[t!]
\begin{center}
\hspace*{-2mm}
\includegraphics[scale=0.61]{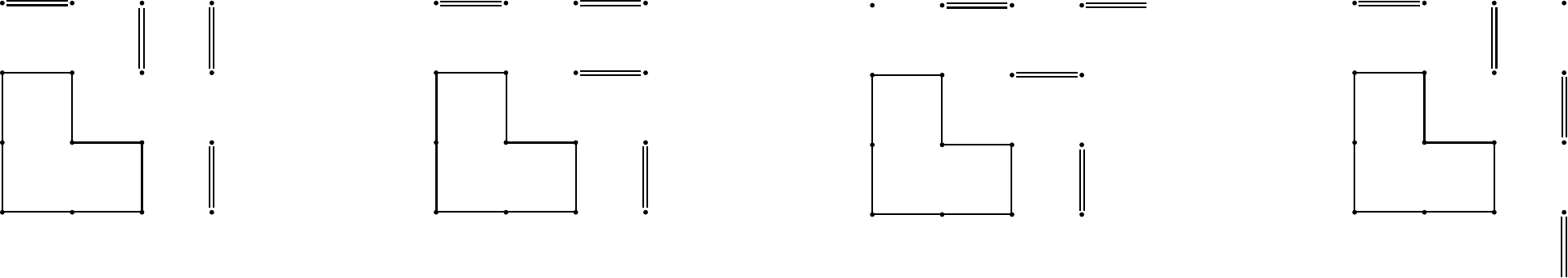}
\vspace{8mm}

\includegraphics[scale=0.61]{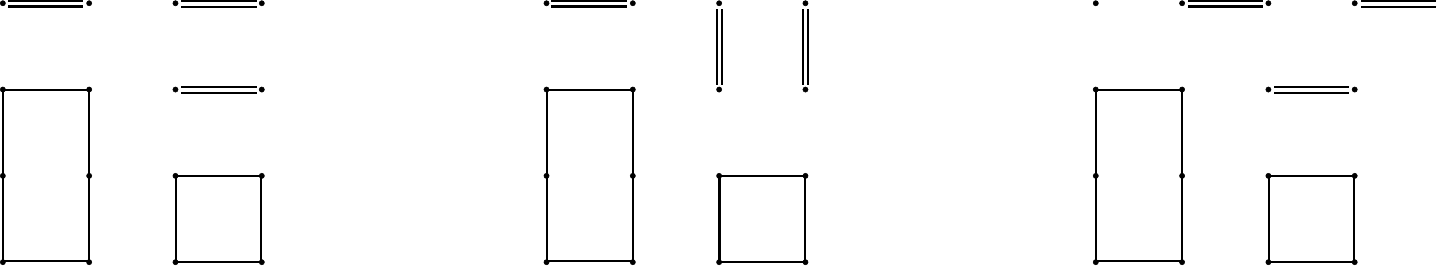}
\vspace{3mm}

\caption{Examples of possible fermion loop configurations and all dimer assignments compatible with a given loop 
configuration (a single fermion loop in the example in the top row, and two separate loops in the second example (bottom)). The lattice 
size is $4 \times 4$ with periodic boundary conditions, and the loops are shown without orientation.} 
\label{countexamples}
\end{center}
\end{figure}

In this appendix we briefly describe the exact summation of the dual partition sum on a small lattice ($4 \times 4$ in our case),
starting the discussion with the one flavor case. The first observation is that the number of admissible configurations of fermion 
loops and dimers (every site of the lattice is either run through by a loop or is the endpoint of a dimer) is finite. 
On a sufficiently small lattice all admissible configurations can be generated in a computer 
program which first places loops on the lattice and then fills the remaining sites with dimers (see the examples in 
Fig.~\ref{countexamples}). Of course loops that are not compatible with any valid dimer configuration have to be excluded, such as,
e.g., a loop around $2 \times 2$ plaquettes with an isolated single site inside that cannot be connected to another site with a dimer. Also 
configurations with a non-zero net winding number are excluded, since they cannot be saturated with occupied plaquettes
(compare the discussion of Gauss' law in Section 2). 

In most cases a given loop configuration is compatible with several dimer configurations. In the example in the top row of 
Fig.~\ref{countexamples} there are 4 ways for placing the dimers, in the bottom example there are three possibilities. 
Since the dimers do not affect the weigths in the 
dual representation (\ref{zfinal2}) the number of dimer configurations compatible with a given set of loops enters as a degeneracy factor.
This factor is 4 in the example in the top row of Fig.~\ref{countexamples} and 3 in the bottom. 

The next step is to analyze the plaquette occupation numbers that are compatible with the given loop configurations. For this the possible
orientations of the loops have to be taken into account. To illustrate the procedure we start with the top row example in 
Fig.~\ref{countexamples} and assume that the loop is oriented in the mathematically negative sense. A possible configuration of plaquette
occupation numbers then is given be $p(n) = 0$ for the 13 plaquettes outside the loop and $p(n) = +1$ for the 3 plaquettes inside
the loop (lattice size is $4 \times 4 = 16$). 
However, more generally one can set $p(n) = k$ for the 13 plaquettes outside the loop and $p(n) = k + 1$ for the plaquettes 
inside, where $k$ is an arbitrary integer. The complete contribution then is obtained by summing over all possible $k \in \mathds{Z}$. 
In case the orientation of the loop is chosen in the mathematically positive sense (both orientations have to be summed), the admissible
plaquette configurations are $p(n) = k$ for the plaquettes outside the loop and $p(n) = k - 1$ inside. Thus the full contribution of the 
diagram in the top row of Fig.~\ref{countexamples} to the partition sum (\ref{zfinal2}) is given by
\begin{equation}
4 \sum_{k \in \mathds{Z}} \!
\left(\! I_{|k|}\!\left( 2 \sqrt{\eta \overline{\eta}} \, \right)  \left( \sqrt{\frac{\eta}{\overline{\eta}}} \, \right)^{\!k} \right)^{13} 
\!\! \times \! 
\left[ \left( \!I_{|k+1|}\!\left( 2 \sqrt{\eta \overline{\eta}} \, \right) \left( \sqrt{\frac{\eta}{\overline{\eta}}} \, \right)^{\! k+1} \right)^{\!3} \!
+ \left( \!I_{|k-1|}\!\left( 2 \sqrt{\eta \overline{\eta}} \, \right) \left( \sqrt{\frac{\eta}{\overline{\eta}}} \, \right)^{\! k-1} \right)^{\!3\;} \right] .
\end{equation}
The asymptotic behavior $I_k(x)  = I_{-k}(x) \sim ( e \, x / 2 k)^k / \sqrt{2 \pi k}$ of the modified
Bessel functions for $k \rightarrow \infty$ (see, e.g., \cite{nist}) guarantees the (fast) convergence of the sum over $k$. 

For the example in the bottom plot of Fig.~\ref{countexamples} we have to sum over all four possible combinations of orientations of the 
two loops. An easy generalization of the first example then gives the corresponding contribution
\begin{eqnarray}
\;&& \hspace*{-8mm} 3 \sum_{k \in \mathds{Z}} \!
\left(\! I_{|k|}\!\left( 2 \sqrt{\eta \overline{\eta}} \, \right)  \left( \sqrt{\frac{\eta}{\overline{\eta}}} \, \right)^{\!k} \right)^{13} 
\!\! \times \! 
\Bigg[ \left( \!I_{|k+1|}\!\left( 2 \sqrt{\eta \overline{\eta}} \, \right) \left( \sqrt{\frac{\eta}{\overline{\eta}}} \, \right)^{\! k+1} \right)^{\!3} \!
+ \left( \!I_{|k-1|}\!\left( 2 \sqrt{\eta \overline{\eta}} \, \right) \left( \sqrt{\frac{\eta}{\overline{\eta}}} \, \right)^{\! k-1} \right)^{\!3} 
\\
\; && \hspace{2mm} +
\left(I_{|k+1|}\!\left( 2 \sqrt{\eta \overline{\eta}} \, \right)\right)^2 \, I_{|k-1|}\!\left( 2 \sqrt{\eta \overline{\eta}} \, \right)
\left( \sqrt{\frac{\eta}{\overline{\eta}}} \, \right)^{\! 3k+1} +
\left(I_{|k-1|}\!\left( 2 \sqrt{\eta \overline{\eta}} \, \right)\right)^2 \, I_{|k+1|}\!\left( 2 \sqrt{\eta \overline{\eta}} \, \right)
\left( \sqrt{\frac{\eta}{\overline{\eta}}} \, \right)^{\! 3k-1}
\Bigg] . \nonumber 
\end{eqnarray}  
Similar to the two examples discussed here, one can generate the sums over all possible loop configurations in a 
simple computer program, together with the corresponding degeneracy factors from the number of dimer configurations 
compatible with a given loop configuration. 

The exact summation program can be generalized to the two flavor case in a straightforward way. The fermion constraints have to hold
for both flavors independently, such that one obtains the set of admissible two flavor fermion configurations as the product of the 
sets of admissible fermion configurations for both flavors. The corresponding plaquette occupation numbers can be determined similarly
to the one flavor case, with the small modification that for saturating the gauge constraints for the second flavor the plaquette flux 
or the orientation of a loop flux from the first flavor have to run in the same direction of the flux of the second flavor since it has 
opposite charge. For that reason, in the two flavor case also configurations where loops of the two flavors wind the same number of 
times around the periodic boundaries are possible. In case of a temporal winding they contribute to the dependence on the chemical potentials, i.e., their contributions are weighted with $e^{N_T (\mu_\psi + \mu_\chi) \, W}$, where $W$ is the temporal 
winding number ($W$ must be the same for both flavors).

\end{document}